\documentclass[10pt,twoside,twocolumn,superscriptaddress,aps,prx%
,reprint,final,longbibliography
]{revtex4-2}
\usepackage[utf8]{inputenc}
\usepackage[T1]{fontenc}

\usepackage{graphicx,latexsym} 
\graphicspath{{images/}{../images/}}
\usepackage{amssymb,amsthm,amsmath,amsfonts}
\usepackage{bm}
\usepackage{dcolumn}
\usepackage{gensymb}
\usepackage{braket}
\usepackage{siunitx}
\usepackage{layouts}
\usepackage{lipsum}
\usepackage{float}
\usepackage{braket}

\newcommand{\sgn}[1]{\mathcal{S}(#1)}

\usepackage[]{xcolor}
\usepackage{hyperref}
\hypersetup{
	colorlinks,breaklinks,
	linkcolor={blue!90!black},
	citecolor={blue!90!black},
    urlcolor={blue!90!black},
}
\usepackage[]{newtxtext}
\usepackage[subscriptcorrection,nosymbolsc,smallerops,bigdelims]{newtxmath}
\DeclareMathAlphabet{\mathcal}{OMS}{cmsy}{m}{n}
\DeclareMathAlphabet{\mathbcal}{OMS}{cmsy}{b}{n}

\newcommand{\AffCam}{Cavendish Laboratory, University of Cambridge, J.J. Thomson Avenue, Cambridge, CB3 0HE, UK}
\newcommand{\AffLinz}{Institute of Semiconductor and Solid State Physics, Altenberger Str. 69, Johannes Kepler University, Linz, Austria}
\newcommand{\AffWroclaw}{Institute of Theoretical Physics, Wroc{\l}aw University of Science and Technology, Wroc{\l}aw, Poland}
\newcommand{\AffQuandela}{Quandela, 7 rue Leonard da Vinci, 91300, Massy, France}
\newcommand{\AffQParis}{Centre for Nanosciences and Nanotechnologies, CNRS, Université Paris-Saclay, UMR 9001, 10 Boulevard Thomas Gobert, 91120, Palaiseau, France}
\newcommand{\AffBasel}{Department of Physics, University of Basel, Klingelbergstrasse 82, 4056 Basel, Switzerland}
\newcommand{\AffOxford}{Department of Engineering Science, University of Oxford, Parks Road, OX1 3PJ, UK }
\newcommand{\AffCampinas}{Instituto de Física Gleb Wataghin, Universidade Estadual de Campinas, 13083-859 Campinas, Brazil}
\newcommand{\AffBochum}{Lehrstuhl für Angewandte Festkörperphysik, Ruhr-Universität Bochum, 44780 Bochum, Germany}

\begin{document}
\preprint{APS/123-QED}

\title{Optical and magnetic response by design in GaAs quantum dots}

\author{Christian Schimpf}
\email[Corresponding author: ]{cs2273@cam.ac.uk}
\affiliation{\AffCam}
\author{Ailton J. Garcia Jr.}
\affiliation{\AffLinz}
\author{Zhe X. Koong}
\affiliation{\AffCam}
\author{Giang N. Nguyen}
\affiliation{\AffBasel}
\author{Lukas L. Niekamp}
\affiliation{\AffBasel}
\author{Martin Hayhurst Appel}
\affiliation{\AffCam}
\author{Ahmed Hassanen}
\affiliation{\AffCam}
\affiliation{\AffOxford}
\author{James Waller}
\affiliation{\AffCam}
\author{Yusuf Karli}
\affiliation{\AffCam}
\author{Saimon Filipe Covre da Silva}
\affiliation{\AffLinz}
\affiliation{\AffCampinas}
\author{Julian Ritzmann}
\affiliation{\AffBochum}
\author{Hans-Georg Babin}
\affiliation{\AffBochum}
\author{Andreas D. Wieck}
\affiliation{\AffBochum}
\author{Anton Pishchagin}
\affiliation{\AffQuandela}
\author{Nico Margaria}
\affiliation{\AffQuandela}
\author{Ti-Huong Au}
\affiliation{\AffQuandela}
\author{Sebastien Boissier}
\affiliation{\AffQuandela}
\author{Martina Morassi}
\affiliation{\AffQParis}
\author{Aristide Lemaitre}
\affiliation{\AffQParis}
\author{Pascale Senellart}
\affiliation{\AffQParis}
\author{Niccolo Somaschi}
\affiliation{\AffQuandela}
\author{Arne Ludwig}
\affiliation{\AffBochum}
\author{Richard J. Warburton}
\affiliation{\AffBasel}
\author{Mete Atat\"ure}
\affiliation{\AffCam}
\author{Armando Rastelli}
\affiliation{\AffLinz}
\author{Micha{\l} Gawe{\l}czyk}
\email[Corresponding author: ]{michal.gawelczyk@pwr.edu.pl}
\affiliation{\AffWroclaw}
\author{Dorian A. Gangloff}
\email[Corresponding author: ]{dag50@cam.ac.uk}
\affiliation{\AffCam}

\begin{abstract}
Quantum networking technologies use spin qubits and their interface to single photons as core components of a network node. This necessitates the ability to co-design the magnetic- and optical-dipole response of a quantum system. These properties are notoriously difficult to design in many solid-state systems, where spin-orbit coupling and the crystalline environment for each qubit create inhomogeneity of electronic $g$-factors and optically active states. Here, we show that GaAs quantum dots (QDs) obtained via the quasi-strain-free local droplet etching epitaxy growth method provide spin and optical properties predictable from assuming the highest possible QD symmetry. Our measurements of electron and hole $g$-tensors and of transition dipole moment orientations for charged excitons agree with our predictions from a multiband $\bm{k}\cdot\bm{p}$ simulation constrained only by a single atomic-force-microscopy reconstruction of QD morphology. This agreement is verified across multiple wavelength-specific growth runs at different facilities within the range of 730~nm to 790~nm for the exciton emission. Remarkably, our measurements and simulations track the in-plane electron $g$-factors through a zero-crossing from $-0.1$ to $0.3$ and linear optical dipole moment orientations fully determined by an external magnetic field. The robustness of our results demonstrates the capability to design -- prior to growth -- the properties of a spin qubit and its tunable optical interface best adapted to a target magnetic and photonic environment with direct application for high-quality spin-photon entanglement.

\end{abstract}

\maketitle


\section{Introduction}

Entanglement between matter-based and light-based qubits is the key resource in quantum communication and distributed quantum computation \cite{Briegel2009,Awschalom2018}, and to deterministically generate resource states for measurement- and fusion-based quantum computation \cite{Briegel1998,Raussendorf2003,Azuma2015,Ruf2021}. Amongst the variety of physical platforms being explored for this task, including atom-cavity systems \cite{Thomas2022}, solid-state structures hold promise for compact and integrated systems that can be deployed at scale for useful applications. These include diamond or SiC color centers \cite{Stas2022, Bourassa2020}, rare-earth ions \cite{Ruskuc2021a}, and semiconductor quantum dots (QDs) \cite{Tomm2021, Appel2025} as promising candidates that satisfy the basic requirements of marrying a coherent spin-photon interface with a long-lived quantum memory. However, solid-state systems present unique challenges, including phonon coupling \cite{Reiter2019}, charge noise, and nuclear-spin noise \cite{Kuhlmann2013}, and have so far lacked the precise control and reproducibility needed for scalable quantum devices.

\begin{figure*}
    \centering
    \includegraphics[scale = 0.95]{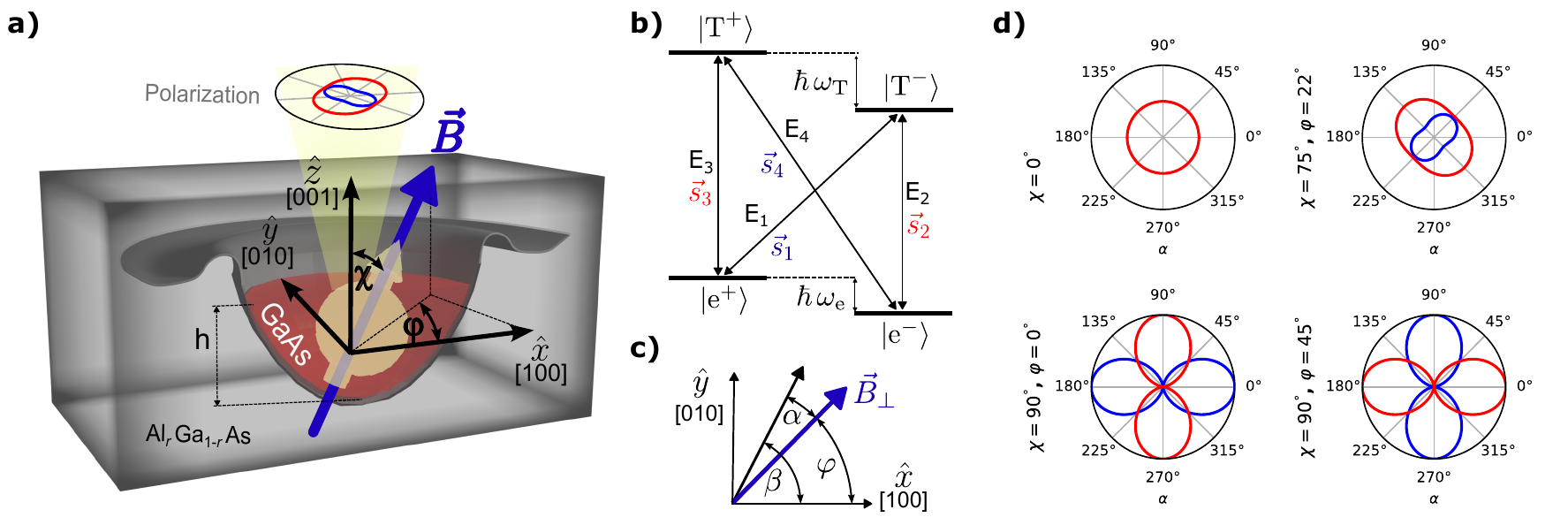}
    \caption{(a) Schematic of a spin-photon interface in a GaAs QD obtained by local Al droplet etching. A single electron is confined in the QD consisting of pure GaAs with a filling height of $h$, buried in an $\text{Al}_r\text{Ga}_{1-r}\text{As}$ host matrix. (b) Energy-level structure of a GaAs QD in a static magnetic field $\bm{B}$ at a polar angle $\chi$ to the [001] crystal axis ($\hat{z}$) (the growth- and optical axis) and an azimuthal angle $\varphi$ relative to $[100]$ ($\hat{x}$). The ground-state electron and the negative trion split into the states $\ket{e^{\pm}}$ and $\ket{T^{\pm}}$, respectively. The four possible optical transitions $\text{E}_{1-4}$ with their Stokes vectors $\bm{s}_{1-4}$ are ordered by their energies in ascending order. (c) In-plane magnetic field component $\bm{B}_\perp$ aligned at an angle $\varphi$ relative to $\hat{x}$. An arbitrary angle of interest is either given as $\alpha$ relative to $\bm{B}_\perp$ or as $\beta$ relative to $\hat{x}$. (d) Polar plots of the linear projection of the optical transition dipole moment for transitions $E_{1,4}$ (blue) and $E_{2,3}$ (red) for a system with $D_{2d}$ symmetry and all positive $g$-factors versus $\alpha$ for different configurations of $\chi$ and $\varphi$. The angle $\alpha=\SI{0}{\degree}$ corresponds to the direction of $\bm{B}_\perp$.}
    \label{fig:GaAs}
\end{figure*}

Despite these challenges, semiconductor quantum dots have emerged as commercially viable quantum light sources, owing to their near-unity quantum efficiency, short radiative lifetimes, and high single-photon purity and photon indistinguishability \cite{Rusong2023}. Their light-matter coupling can be further enhanced via photonic structures such as integrated vertical cavities \cite{Somaschi2016}, circular Bragg resonators \cite{Liu2019,Wang2019,Hekmati2023}, photonic crystals \cite{Lodahl2015} or open cavities \cite{Tomm2021}. For the latter, an impressive $>\SI{70}{\percent}$ photon collection efficiency into a single-mode fiber has now been demonstrated \cite{Ding2025}. Recently, chiral interfaces between emitters and integrated photonic circuits were shown to benefit strongly from precise engineering of the photon polarization modes \cite{Lang2022,Rosinski2024}. Going further, an exact alignment of the QD transition dipole moment to its photonic environment will be crucial.

On the spin side, proof-of-concept experiments with electron-, hole- and nuclear spin qubits in QDs \cite{Urbaszek2013,Stanek2014,Gangloff2019,Chekhovich2020,Appel2022,Cogan2023} have highlighted their further potential for single-spin and spin-ensemble physics, with a prospect towards practical quantum memories \cite{Denning2019}, quantum computation \cite{Loss1997,Buterakos2017} and the generation of multi-dimensional photonic cluster states \cite{Michaels2021}. The spin qubit in a QD is formed by splitting the discrete energy levels of a ground-state particle, often a single electron, via a static magnetic field. Spin control and read-out can be realized very efficiently by exploiting their coherent spin-selective optical transitions between a ground state spin and a charged exciton state (``trion'') \cite{Press2008,NickVamivakas2009,Bodey2019,Antoniadis2023}.

Considering the vital role of light-matter-coupling in qubit control and photon extraction, it becomes clear that creating QD devices for integration within scalable photonic quantum technologies requires designing the spin-photon interface \textit{a priori}, i.e., before device fabrication and integration. To do this, we require precise knowledge and control over the energy level structure of the spin qubit and the exciton states, via their $g$-tensor and the radiative transition dipole moments (TDMs) between them. 

The spin qubit and transition dipoles are governed by the magnetic response of the particles involved, described by their effective $g$-tensors and the phase-relations between their wavefunctions \cite{Pikus1994,Semenov2003}. In semiconductor QDs, the $g$-tensors are governed by a competition between the free-particle magnetic response and spin-correlated orbital currents (SCOCs) \cite{VanBree2014} (see Sec. \ref{sec:theory}) resulting from spin-orbit interaction. Analogous to the classical magnetic moment of a current loop, the SCOCs depend on the size and the properties of the system in which they can ``flow''. This leads to a strong sensitivity of the $g$-tensors and the TDMs to QD geometry, material composition, and strain \cite{Koudinov2004, Semenov2003,Kiessling2006,Trifonov2021}, making the $g$-tensors notoriously difficult to predict and to reproduce. In InGaAs, the QD material system most studied to date for both photonic \cite{Vural2020} and spin applications \cite{Wust2016,Stockill2016,Gangloff2019,Jackson2022}, the strain-driven growth leads to a deviation from a radially symmetric shape and alloy composition around the growth axis and to built-in strain. The reduced symmetry leads to complicated dependencies of the TDMs on the external magnetic field \cite{Pikus1994,Semenov2003,Trifonov2021}. For substantial geometric imperfections or large strain, the TDMs are entirely pinned to a specific direction, independent of the magnetic field orientation \cite{VanBree2016, ramesh2025arXiv}, making alignment of TDMs to photonic structures very challenging \cite{Appel2022}. Variations of shape and strain from QD to QD render the $g$-tensors and TDMs of a single InGaAs QD nearly impossible to predict and even difficult to fully determine experimentally \cite{VanBree2016, Kiessling2006}. 

In contrast to that, GaAs QDs obtained via the local Al droplet etching (LDE) technique \cite{Gurioli2019,DaSilva2021}, have demonstrated remarkable reproducibility and homogeneity, which led to the reported low variance in emission wavelengths, low fine structure splitting \cite{Huo2013FSS,DaSilva2021} and Fourier-limited linewidth \cite{Zhai2020}. Recently, electron spin coherence times beyond \SI{100}{\micro s} were demonstrated in such LDE GaAs QDs, outperforming other QD material systems by orders of magnitude and hinting towards nuclear-spin-memory times beyond \SI{100}{ms} \cite{Zaporski2023,Nguyen2023,Appel2025, Dyte2025}. This prior work on LDE GaAs QDs highlights their potential for modern quantum applications that require well-defined spin-photon interfaces and long spin coherence time. Despite impressive advances, the ability to predict and control the optical and magnetic dipole moments of GaAs QDs relative to their photonic and magnetic environments remains a key challenge.

In this work, we show that GaAs QDs adhere closely to the behavior dictated by $D_{2d}$ point group symmetry, making their optical and magnetic dipoles predictable and reliably controllable. To do so, we elucidate, both theoretically and experimentally, the material parameters, mechanisms, and symmetries governing the magnetic dipoles of the electron and the negatively charged trion in LDE GaAs QDs and the optical transition dipoles between them. Using only atomic-force-microscopy (AFM) measurements of the droplet-etched nanoholes \cite{DaSilva2021} as an input, we simulate the full $g$-tensors and TDMs in a framework of combined multiband $\bm{k}\cdot\bm{p}$ \cite{Bahder1992,Gawarecki2014} and configuration-interaction \cite{Bryant1987} methods over a large parameter space. Using resonant spectroscopy and hyperfine-mediated measurements to fully characterize the magnetic and optical response of a QD, we obtain the $g$-tensors of the electron and the trion -- including their sign -- and the involved optical transition dipoles. We demonstrate excellent agreement with our simulations for LDE GaAs QDs from different molecular beam epitaxy growth runs and even different growth facilities over emission wavelengths from \SI{730}{nm} to \SI{790}{nm}, including those with an electron $g$-factor close to zero. Our work confirms that in modern LDE GaAs QDs the high in-plane symmetry, which leads to low fine structure splitting \cite{Huo2013FSS}, also causes the TDMs to strongly adhere to those expected for pure $D_{2d}$ symmetry. This makes the TDMs not only predictable but also fully tuneable by the applied magnetic field. Our work shows that, thanks to predictive theory and extremely reproducible fabrication, designing the spin-photon interface of QD devices integrated with photonic structures is indeed possible and is now at our disposal.

\section{Spin-photon interface in LDE GaAs quantum dots}
\label{sec:theory}

Figure \ref{fig:GaAs}(a) sketches an LDE GaAs QD, using the coordinate system of the crystal axes $\hat{x}$, $\hat{y}$, $\hat{z}$ := [100], [010], [001]. A static magnetic field $\bm{B}$, applied at an angle $\chi$ relative to the optical axis $\hat{z}$ and an angle $\varphi$ to $\hat{x}$, splits the spin of a confined electron into the two Zeeman-eigenstates $\ket{e^{\pm}}$. These states couple to the two negatively charged exciton (trion) states $\ket{T^{\pm}}$ via four possible optical transitions, E$_i$, $i=1,2,3,4$, with their respective TDMs described by the Stokes vectors $\bm{s}_{i}$. The electron has two spin projections $\pm 1/2$, while the trion is composed of two electrons in a spin-singlet configuration and a heavy hole (HH) with two spin projections $\pm 3/2$. The resulting optically active double-lambda system, shown in Fig.~\ref{fig:GaAs}(b), is the workhorse for initializing, reading out, and controlling the electron spin-qubit in III-V semiconductor QDs \cite{Bodey2019}.
For spin control and readout, two extremal configurations of $\chi$ are typically used, which are complementary in their use cases. In the \textit{Faraday} configuration ($\chi=0$), the forbidden diagonal transitions $E_{1,4}$ lead to a high cyclicity of the remaining transitions $E_{2,3}$, which enables non-destructive and single-shot readout \cite{Antoniadis2023}. In the \textit{Voigt} configuration ($\chi=\pi/2$), all four transitions are equally allowed and enable coherent optical control of the spin-qubit via a two-color Raman scheme \cite{Bodey2019}, but readout leads to an almost immediate destruction of the spin state (in an isotropic photonic environment). At oblique angles $0< \chi<\pi/2$ all transitions become partially allowed \cite{Barr2024}.

\begin{figure*}
    \centering
    \includegraphics[scale = 1]{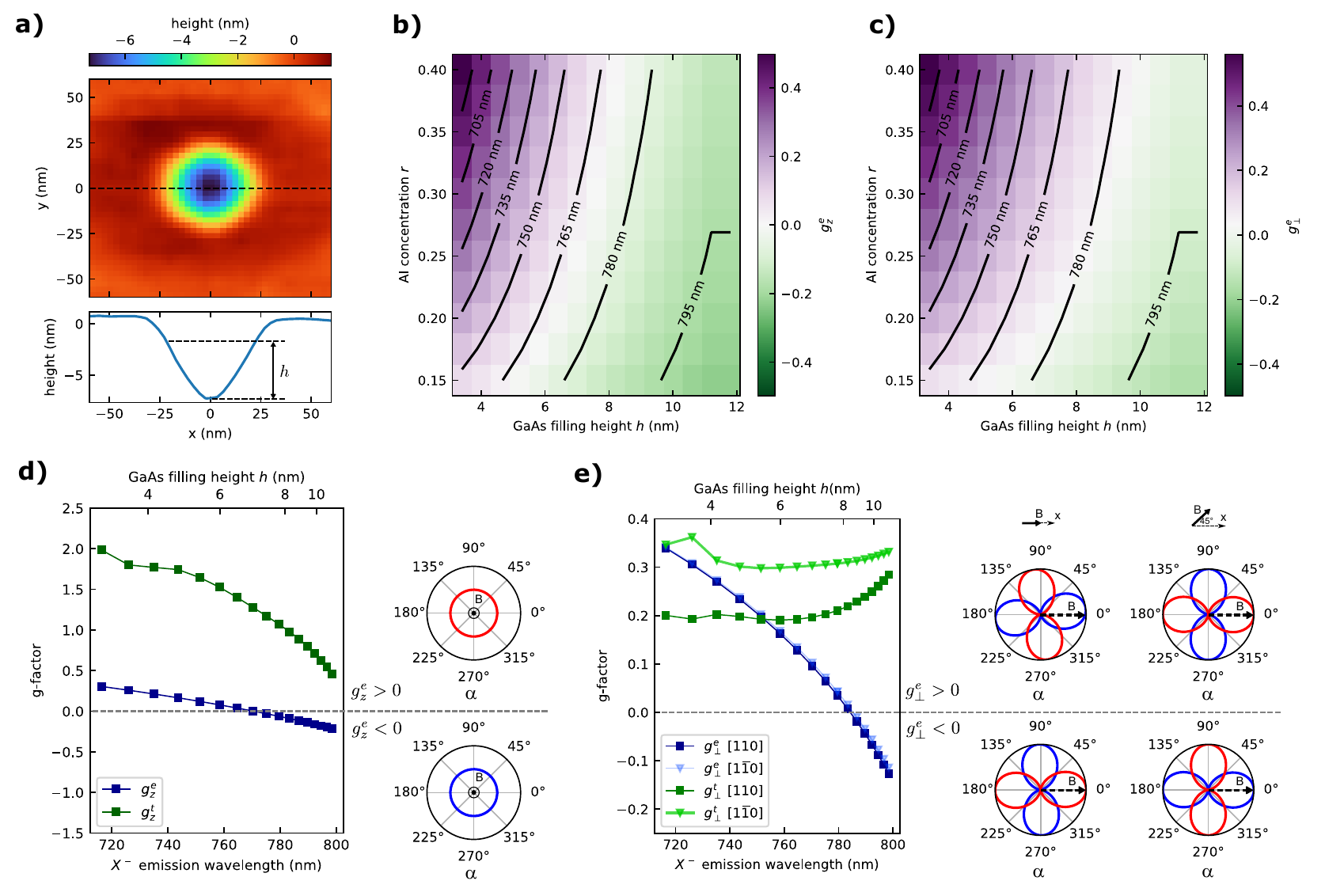}
    \caption{(a) AFM image of an Al droplet etched nanohole, used to form a GaAs QD. The bottom graph shows a cross-section at $y=0$. (b,c) Simulated electron $g$-factors for $\bm{B}$-fields in (b) [001]  and (c) [110] directions as a function of GaAs filling height $h$ and Al$_r$Ga$_{1-r}$As barrier Al concentration $r$. The black lines indicate constant trion-to-electron ($X^-$) emission wavelength. (d,e) Simulated electron- and trion $g$-factors in a GaAs QD in an $\text{Al}_{0.25}\text{Ga}_{0.75}\text{As}$ barrier in the (d) Faraday and (e) Voigt configurations as a function of the $X^-$ emission wavelength. The polar plots show the associated transition dipole moment polarizations $\bm{s}_{1,4}$ (blue) and $\bm{s}_{2,3}$ (red).} 
    \label{fig:simulation}
\end{figure*}

The high rotational symmetry of the cone-shaped LDE GaAs QDs around the $\hat{z}$-axis and the low built-in strain \cite{DaSilva2021} allow us to assume the $D_{2d}$ point symmetry group as a first approximation. Although a conical shape, as opposed to a circular disk, formally reduces the symmetry to $C_{2v}$, this is limited to weak interfacial effects that, as we show, can be neglected. In this limit, we can neglect heavy hole-light hole (HH-LH) mixing. Given this assumption, $\ket{\text{e}^\pm}$ and $\ket{\text{T}^\pm}$ are separated by their respective Zeeman splittings, 
\begin{equation}
\hbar\omega_{\text{e,t}} = \mu_B\,|\bm{B}|\, \sqrt{ (g^{\text{e,t}}_z \cos{\chi})^2 + (g^{\text{e,t}}_{\perp} \sin{\chi})^2  }, 
\label{eq:omega}
\end{equation}
with $\mu_{\text{B}}$ the Bohr magneton, $g^{\text{e,t}}_z$ the out-of-plane $g$-factors and $g^{\text{e,t}}_{\perp}$ the isotropic in-plane $g$-factors for the electron (e) and the trion (t), respectively. The four emerging transition energies are then given by $E_{1,2}/\hbar=\omega^\prime\,+\, (-\omega_t \mp \omega_e)/2$ and $E_{3,4}/\hbar=\omega^\prime\,+\,(\omega_t \mp \omega_e)/2$, with $\omega^\prime$ the central transition frequency.

The magnetic response of an isolated particle in a vacuum is described by the Land\'e $g$-factor,
\begin{equation}
    g_l = 2\,m_J\left(1 + \frac{J(J+1) + S(S+1) - L(L+1)}{2J(J+1)}\right),
\label{eq:Lande}
\end{equation}
such that, in analogy to Eq. \eqref{eq:omega}, the total energy splitting in a magnetic field is $\mu_{\text{B}}\,|\bm{B}\,|g_l$ (i.e., we incorporate $m_J$ into $g_l$ to adhere to common experimental nomenclature), with $L$ the orbital angular momentum, $S$ the spin, $J=L+S$ the total angular momentum and $m_J$ the projection of $J$ along $\bm{B}$. For an electron in a III-V semiconductor conduction band (CB), $S=1/2$, $L=0$, and $m_J=1/2$ this results in $g_l=2$, while for a valence band HH, $S=1/2$, $L=1$, and $m_J=3/2$ this results in $g_l=4$. In a bulk semiconductor with nonzero spin-orbit coupling, such as GaAs, this single-particle magnetic momentum has an added contribution from induced spin-correlated orbital currents (SCOCs) perpendicular to $\bm{B}$ \cite{VanBree2014}. The SCOCs depend on both the particle's envelope function and the lattice-periodic Bloch function \cite{Chuang2009}. For the electron, with a spherically symmetric Bloch function, the total magnetic response converges to the isotropic $\Gamma$-point bulk $g$-factor described by Roth's formula \cite{Roth1959},
\begin{equation}
    g_c = 2-\frac{2E_p\Delta_{\text{SO}}}{3E_g\left(E_g+\Delta_{\text{SO}}\right)},
\label{eq:Roth}
\end{equation}
with $E_p$ the interband coupling strength, $E_g$ the bandgap energy and $\Delta_{\text{SO}}$ the valence-band spin-orbit splitting. For bulk GaAs at cryogenic temperatures $E_p=\SI{28.8}{eV}$, $E_g=\SI{1.519}{eV}$ and $\Delta_{\text{SO}}=\SI{0.341}{eV}$ \cite{Vurgaftman2001,Winkler2003}, which results in $g_c= -0.44$, i.e. a negative contribution from the SCOCs dominates the bulk magnetic response.

In a QD, the confinement potential constrains the carrier motion to a region comprised of about $10^5$ lattice sites \cite{Zaporski2023}, which quenches the SCOCs, and hence their contribution to the $g$-factor. This can also be understood as the effect of quenching of electron angular momentum, and thus spin-orbit effects, at higher wavevector components enforced in QD states by spatially finite confinement as opposed to $\Gamma$-point bulk electrons \cite{Gawarecki2020}. We can, therefore, intuitively understand how the reported near-zero electron $g$-factors in LDE GaAs QDs \cite{Zaporski2023,Nguyen2023,Appel2025} can emerge from a balance between the single particle response and the SCOCs. The trion's magnetic response is dominated by the HH, for which the SCOCs are more complex because of the lack of spherical symmetry of the $p$-like Bloch function \cite{Pikus1994,ivchenko2012, VanBree2016}. But they still provide a competing contribution, i.e., one that is opposite to the free-particle magnetic moment. In the Faraday configuration, values of $g^t_z \gg g^e_z$ are reported \cite{Puebla2013}, which means that the HH $g$-factor tends more closely towards the isolated-particle limit than the electron $g$-factor. In the Voigt configuration, the direct Zeeman coupling strength for HHs is zero due to the symmetry of the HH Bloch functions and the growth axis being distinguished. Instead, $g^t_{\perp}$ is governed by non-Zeeman interactions and mixing between HH and light-hole (LH) subbands \cite{Golub1999,Semenov2003,Trifonov2021}, which we will discuss in more detail in Sec. \ref{sec:sim}. 

This qualitative insight into the magnetic response of the pure HHs already allows us to determine the orientation of the TDMs for an ideally $D_{2d}$-symmetric case, where the phase relations between $\ket{e^{\pm}}$ and $\ket{T^{\pm}}$ play a key role (see Supplementary Material for a detailed derivation). The in-plane angles relevant in the following are shown in Fig. \ref{fig:GaAs}(c), with $\varphi$ the angle between the in-plane component of the magnetic field $\bm{B_{\perp}}$ and $\hat{x}$, $\alpha$ the angle between an arbitrary vector of interest and $\bm{B_{\perp}}$, and $\beta=\varphi+\alpha$. Figure \ref{fig:GaAs}(d) then depicts the resulting TDM polarizations for four different orientations of $\bm{B}$. In the Faraday configuration ($\chi=0$), the angular momentum along the $\hat{z}$-axis is conserved due to the radial symmetry so that only $E_2$ and $E_3$ are dipole-allowed if $\sgn{g^{\text{e}}_z}=\sgn{g^{\text{t}}_z}$, or only $E_1$ and $E_4$ if $\sgn{g^{\text{e}}_z}\neq\sgn{g^{\text{t}}_z}$, where $\sgn{x}$ denotes the sign. In both cases, the polarizations are left- and right-circular, respectively \cite{Barr2024}. In the Voigt configuration ($\chi=\pi/2$), $E_{1-4}$ are equally dipole-allowed, where $\bm{s}_1=\bm{s}_4$ are linear and orthogonal to $\bm{s}_2=\bm{s}_3$. The angles $\beta_{1,2}$ between $\bm{s}_{1,2}$ and $\hat{x}$ become \cite{Pikus1994,Kiessling2006,Trifonov2021}
\begin{equation}
  \begin{aligned}
  \beta_1 = -\varphi ,\ \beta_2 = -\varphi + \frac{\pi}{2} \quad & \text{for }\sgn{g^{\text{e}}_{\perp}}=\sgn{g^{\text{t}}_{\perp}},\\    
  \beta_1 = -\varphi + \frac{\pi}{2},\ \beta_2 = -\varphi \quad& \text{for }\sgn{g^{\text{e}}_{\perp}}\neq\sgn{g^{\text{t}}_{\perp}}.
  \end{aligned}
\label{eq:relAngles}
\end{equation}
Interestingly, we can observe from Eq. \eqref{eq:relAngles} that $\bm{s}_1$ or $\bm{s}_2$ align with $\bm{B}$ for each turn of $\varphi$ by $\pi/4$. As shown in Fig.~\ref{fig:GaAs}(d), for $\varphi=0$ the magnetic field $\bm{B}$ is aligned with $\hat{x}$ and $\bm{s}_{1,4}$ are aligned with $\bm{B}$ (in the case of $\sgn{g^{\text{e}}_{\perp}}=\sgn{g^{\text{t}}_{\perp}}$). For other angles, the TDMs counter-rotate with respect to $\bm{B}$. For example, rotating $\bm{B}$ to the [110] crystallographic direction at $\varphi=\pi/4$ rotates the linear polarization by $-\pi/4$, so that $\bm{s}_{2,3}$ instead are aligned with $\bm{B}$. Note that the behavior in Eq.~\eqref{eq:relAngles} can only hold for highly in-plane symmetric QDs with (atomistically) homogeneous QD material and negligible interfacial effects. Deformation, strain, material alloying in the QD, or defects can lead to strikingly different results, as clearly observed in InGaAs QDs \cite{Semenov2003,Trifonov2021,ramesh2025arXiv}. For angles $0<\chi<\pi/2$, the polarization becomes a superposition between the Faraday and the Voigt cases (see Supplementary Material for details).

\section{Simulation from atomic force microscopy data}
\label{sec:sim}

The assumption of a global $D_{2d}$ symmetry holds strictly for GaAs/AlGaAs quantum wells, where the GaAs layer extends out quasi-infinitely in the $\hat{x}$ and $\hat{y}$ directions. However, in the case of LDE GaAs QDs, even though the overall geometry has almost ideal rotational symmetry, the inclined sidewalls of the cone-like geometry combined with the tetrahedral crystal symmetry formally result in the $C_{2v}$ symmetry due to the lack of inversion symmetry in the $\hat{z}$-direction. This effect is expected to be weaker than explicit QD in-plane shape anisotropy, as it exhibits only at the GaAs-AlGaAs interface \cite{Luo2015}. In general, $C_{2v}$ symmetry can lead to HH-LH mixing and anisotropy in the in-plane $g$-factors. To assess the importance of these deviations from the $D_{2d}$ symmetric case discussed in Sec.~\ref{sec:theory} and to calculate the $g$-tensor values, we perform simulations in the multiband $\bm{k}\cdot\bm{p}$ framework \cite{Bahder1992,Gawarecki2014} combined with the configuration-interaction method \cite{Bryant1987} for trion states. As an input, we only take the shape of a nanohole created during the growth process, as shown in the AFM image in Fig. \ref{fig:simulation}(a), and parametrize the nominal GaAs filling height $h$ and the Al fraction $r$ of the $\text{Al}_r\text{Ga}_{1-r}\text{As}$ barrier. This allows us to calculate the wave-function envelopes of both the electron and the hole and further the eigenstates of the trion, including the admixtures of electron and hole excited levels present in the trion ground state. For all these states, we calculate the response to magnetic fields.
Figure \ref{fig:simulation}(b) and (c) show the calculated $g^{\text{e}}_z$ and $g^{\text{t}}_z$, respectively, as a function of $h$ and $r$. The solid curves highlight the constant trion-to-electron ($X^-$) emission wavelengths. The emission wavelength is strongly correlated with both $h$ and $r$, as these two values mostly determine the discrete energy levels in the QD \cite{DaSilva2021}. The two color maps can serve as a guideline for designing QDs to achieve the desired magnetic response. From here on, we will use the $X^-$ emission wavelength, as tuned exclusively by $h$, as our control parameter.

Figure \ref{fig:simulation}(d) shows the Faraday configuration response, i.e., $g^{\text{e}}_z$ and $g^{\text{t}}_z$ as a function of $X^-$ emission wavelength at a typical value of $r=0.25$. For $g^{\text{e}}_z$ we observe a zero-crossing at about \SI{770}{nm}, where the spin-orbit effects begin to dominate the magnetic response. Larger QDs have higher emission wavelengths and also more dominant SCOCs, pushing the $g$-factor towards the bulk limit of $-0.44$. The trend for $g_t^z$ is similar but with overall higher values compared to $g_e^z$. This behavior reflects both the higher bare particle Land\'e $g$-factor value in the valence band and quenched SCOCs for the HH due to its $p$-like Bloch functions, which make the $g$-factor tend more towards the isolated hole value of 4. The polarization of the two allowed transitions for the two scenarios $g^{\text{e}}_z>0$, $g^{\text{t}}_z>0$ (upper right) and $g^{\text{e}}_z<0$, $g^{\text{t}}_z>0$ (lower right), are almost perfectly circular and opposite. These almost perfectly circular polarizations again indicate the behavior expected from a fully $D_{2d}$ symmetric system free of LH effects. 

The behavior of $g$-factors for the Voigt configuration is more complex, as shown in Fig.~\ref{fig:simulation}(e). The dependency of $g^e_\perp$ on the emission wavelength is very similar to the case of $g^e_z$, but the zero-crossing wavelength is offset by about \SI{15}{nm} to \SI{785}{nm}. The offset stems from the quenched SCOCs in the $\hat{z}$-direction (the shortest QD dimension in the plane perpendicular to $\bm{B}$), which decreases the contribution of spin-correlated orbital momentum compared to the Faraday configuration. This quenching pushes the $g$-factor further up towards the free particle limit of 2. The simulation also reveals the existence of anisotropy in $g^{\text{e}}_{\perp}$ and $g^{\text{t}}_{\perp}$, with principal axes along the [110] and the [1$\bar{1}$0] crystallographic directions. These anisotropies originate from the global $C_{2v}$ symmetry when considering the non-equivalence of [110] and [1$\overline{1}$0] directions in (Al)GaAs, in combination with the distinguished direction in the z-axis, and persist even in the absence of strain and for perfectly in-plane symmetric QDs \cite{Luo2015}. Although the absolute anisotropy $\delta g_\perp^e \approx 0.005$ is small, its significant relative value $\delta g_\perp^e/\braket{g_\perp^e}$ in LDE GaAs QDs, with $\braket{g_\perp^e}$ the $g$-factors averaged over both in-plane directions, leads to a tuneable noncollinear electro-nuclear interaction, as observed experimentally \cite{shofer2024ArXiv}. This interaction allows for an electron-spin-dependent feedback loop to be established with nuclear spins, which we exploit in Sec.~\ref{sec:measurement} to determine the sign of the $g$-tensor components. The trion exhibits an almost constant $\braket{g^t_{\perp}}\approx 0.25$ and a significant anisotropy $\delta g^{\text{t}}_{\perp}$ up to about 0.15.

The built-in biaxial strain due to lattice mismatch in QDs splits the HH and LH bands by $\Delta_{\text{LH}}$. That strain in LDE GaAs QDs is very low (about \SI{0.05}{\percent}) but not negligible and results in a minor HH-LH splitting of $\Delta_{\text{LH}}/2\pi\approx\SI{750}{GHz}$ \cite{Huo2013} (two orders of magnitude lower compared to InGaAs QDs). The lack of $\hat{z}$-axis symmetry enables the ``same-spin'' HH-LH mixing effect between HH subband with $m_J=3/2$ and LH with $m_J=1/2$ (and the same for flipped spins), typically referred to as $S$ matrix element in the Luttinger-Kohn Hamiltonian \cite{Bahder1992}. Additionally, any in-plane asymmetry of the overall confining potential enables the ``opposite-spin'' mixing, $\pm3/2\leftrightarrow\mp1/2$ given by the $R$ matrix element. Perturbatively, those effects lead to LH contributions to the hole ground state of magnitudes $\lvert\langle S \rangle/\Delta_{\text{LH}}\rvert^2$ and $\lvert\langle R \rangle/\Delta_{\text{LH}}\rvert^2$, respectively. Since $\Delta_{\text{LH}}$ is low, even weak mixing effects lead to an overall LH contribution of $\approx \SI{7.6}{\percent}$ within the QD volume, according to our simulation. Despite this significant HH-LH mixing, the TDMs are almost perfectly circular in the Faraday configuration (99.7\% degree of circular polarization, see Supplementary Material) and in the Voigt configuration also closely follow the $D_{2d}$ symmetry prediction of Eq.~\eqref{eq:relAngles}. This behavior is evidenced by the $-\pi/2$ rotation of the linear polarization relative to $\bm{B}$ for a $\pi/4$ rotation of $\varphi$, as shown on the right side of Fig.~\ref{fig:simulation}(e). While the ``same-spin'' LH admixture does not add to the same optical transition as the HH, the ``opposite-spin'' admixture can contribute with a polarization orthogonal to the HH's. To understand the robustness of the optical properties despite the presence of the ``opposite-spin'' LH admixture, we have to consider its spatial dependence. Close to the GaAs/AlGaAs interface, the bi-axial strain in the QD crosses zero and becomes compressive in the AlGaAs barrier (see Supplementary Material for details). For in-plane symmetric QDs, this leads to a high HH-LH admixture close to the QD border but a vanishing one at the QD center, where the electron resides. Additionally, the LH admixture envelope is predominantly odd, so its overlap with the electrons is close to vanishing. As a consequence, the HH-LH mixing contributes significantly to the energy shifts of the trion via $g^t$, but the effect on the TDM polarization remains close to negligible. These findings resolve the apparent contradiction between the presence of significant HH-LH mixing \cite{Huber2019} and the near-perfect polarization-entangled photon pairs produced by LDE GaAs QDs \cite{BassoBasset2021,Schimpf2021}. We provide more detailed insight into HH-LH mixing and its constituents in the Supplementary Material.

\section{Optical and spin-polarization measurements}
\label{sec:measurement}

\begin{figure*}
    \centering
    \includegraphics[scale = 0.95]{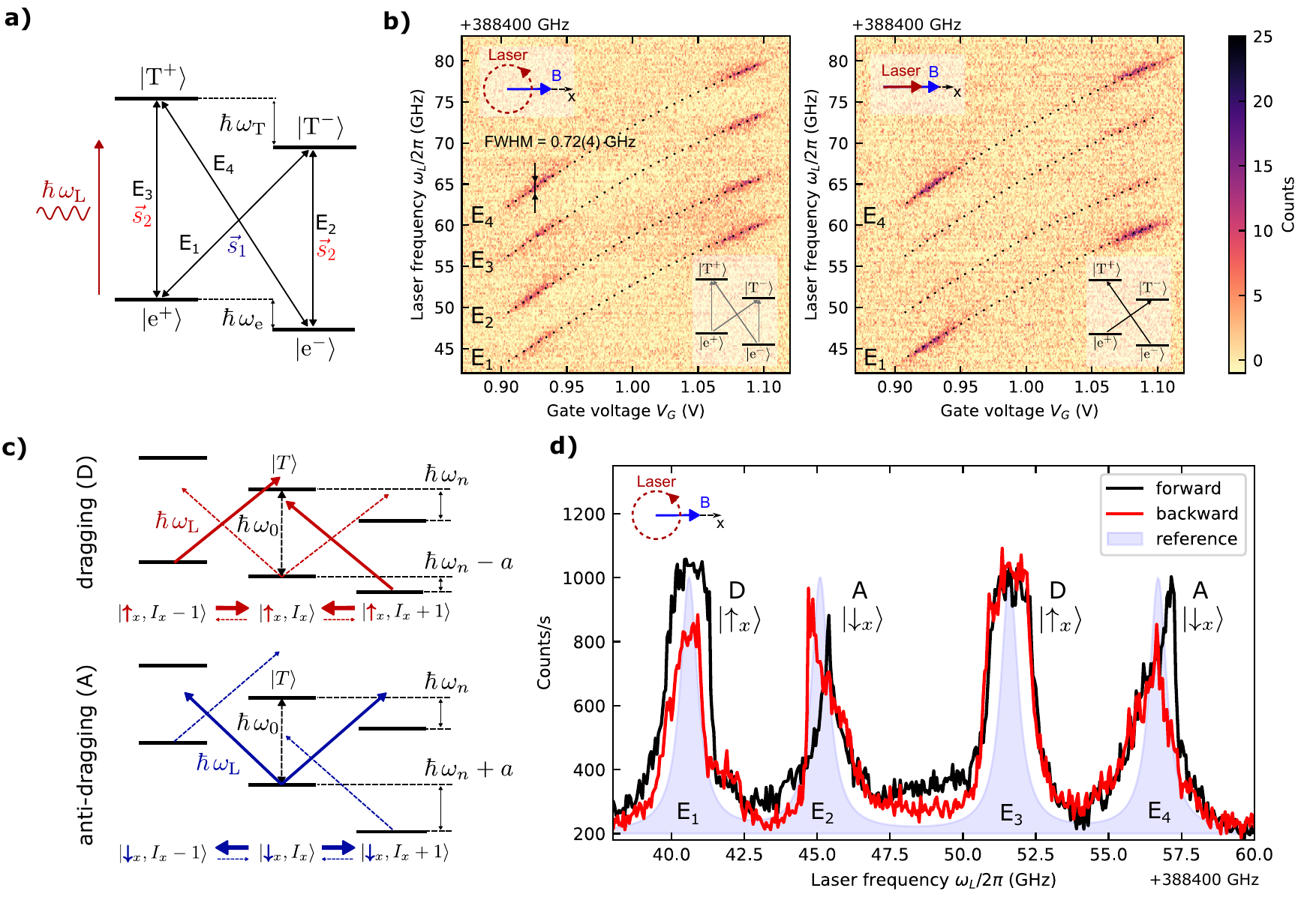}
    \caption{(a) Scheme of the $\ket{T^\pm}$ and $\ket{e^\pm}$ energy levels in the Voigt configuration ($\chi=\pi/2$), driven by a narrow-band CW laser with frequency $\omega_L/2\pi$. In this configuration $\bm{s}_1=\bm{s}_4$ and $\bm{s}_2=\bm{s}_3$. (b) Resonance fluorescence (RF) emission intensity of the trion as a function of the gate voltage $V_G$ and $\omega_L$ with $B_x=\SI{5.7}{T}$. Left: A circularly polarized laser couples equally to $\text{E}_{1-4}$, resulting in four resonances. Right: A laser linearly polarized along $\bm{B}$ results in the predominant excitation of $E_{1,4}$, indicating $s_{1,4}^{(1)}\parallel\bm{B}$ and $s_{2,3}^{(1)}\perp\bm{B}$, with $s^{(1)}$ the rectilinear component of $\bm{s}$. The black dotted lines indicate the approximate resonance conditions for each transition. The transition linewidth extracted from this graph is \SI{0.72(5)}{GHz}. (c) Energy levels of the hyperfine-coupled electron-nuclear manifolds with the electron spin $\ket{\uparrow_x}$ (top) and $\ket{\downarrow_x}$ (bottom) drawn as the ground state. The diagonal arrows indicate electron-trion side-band transitions with a difference in $I_x$ of $\delta I_x\pm1$. In Ga and As the hyperfine constant $a>0$, hence the $\ket{\uparrow}$ manifold tends to self-stabilize (``dragging''), while the $\ket{\downarrow}$ manifold avoids the resonance (``antidragging''). (d) Intensity of the trion emission at $\varphi=\SI{45}{\degree}$ with a circular laser polarization at $V_{G}=\SI{0.93}{V}$ and a slowly varying $\omega_L$ from lower to higher frequencies (black) and reverse (red). The dragging transitions (D) show flat-top behavior, while the anti-dragging transitions (A) reveal a triangle-like shape. As a reference, the light blue filled area shows a theoretical pattern given by four Fourier limited transitions in the absence of dragging.}  
    \label{fig:charact}
\end{figure*}

Equipped with detailed predictions on the $g$-tensors and TDMs of LDE GaAs QDs, we are now in a position to verify them experimentally. All measured QDs are embedded in the intrinsic region of a p-i-n diode, as used in Ref.~\onlinecite{Zaporski2023}, to achieve charge control and to load the QD with a single electron (see Supplementary Material for details). The magnetic response in the Faraday configuration is well studied and thus not the focus of our experiments \cite{Puebla2013,Ulhaq2016,Millington2023}. However, pinning down the exact values and signs of $g^{\text{e}}_{\perp}$ and $g^{\text{t}}_{\perp}$ experimentally has proven difficult in the past, as probing only the electron and trion Zeeman splittings and the polarization of the TDMs leaves the $g$-factor signs ambiguous \cite{VanBree2016, Trifonov2021}. This ambiguity can lead to assignments of $g$-factors to the wrong particle. To overcome this hurdle, we make use of an electron-specific magnetic response to disambiguate the $g$-tensor components: dynamic nuclear polarization (DNP) mediated by the Fermi-contact hyperfine interaction \cite{Urbaszek2013}, in which an electron-spin-dependent nuclear polarization shifts the optical resonance frequencies of the QD. All our measurements, therefore, consist of two steps: first, we perform a polarization-resolved resonance-fluorescence (RF) scan of the transitions $E_{1-4}$ to determine all energy splittings and the TDM orientation; second, we sense the electron-spin-dependent hyperfine-shift so that we can assign the correct $g$-tensor components and signs.  

We will now demonstrate this two-step protocol for one example QD. We apply a magnetic field of \SI{5.8}{T} in the $\hat{x}$ direction, resulting in the double-lambda system shown in Fig.~\ref{fig:charact}(a) with four equally allowed dipole transitions and $\bm{s}_1=\bm{s}_4$ and $\bm{s}_2=\bm{s}_3$. The indicated level splittings in the sketch are only one possible example and not to scale, as we seek to determine the exact state configuration. In our first step, we excite the optical transitions between $\ket{T^\pm}$ and $\ket{e^\pm}$ resonantly using a frequency-tunable laser with a frequency $\omega_L/2\pi$ and a linewidth much smaller than the Fourier-limited linewidth $\Gamma_{\text{F}}/2\pi\approx\SI{700}{MHz}$ \cite{Zhai2020}. Our signal are the photons scattered by the excited trion state and collected through a polarizer oriented to filter out the laser excitation. In the measurements shown in Fig.~\ref{fig:charact}(b) (left side), the laser is circularly polarized and thus couples to all four optical transitions with equal strength. The corresponding four bright features at low and high $V_G$ are observed in the diode bias ranges over which fast electron spin reset through co-tunneling from and into the QD competes with optical spin pumping to maintain a bright steady-state fluorescence. The dark regions between each low-high voltage pair (visualized by black dotted lines) indicate slow co-tunneling and thus efficient optical spin-pumping into an electronic dark state \cite{Dreiser2008}. To avoid the buildup of nuclear spin polarization during the measurement, we ramp $V_G$ with a frequency of \SI{1}{kHz} to periodically enter the co-tunneling regime, effectively scrambling the electron- and the nuclear polarization \cite{Zaporski2023}. The inhomogeneous linewidth extracted from this measurement is $\Gamma/2\pi=\SI{0.72(4)}{GHz}$, which sets the resolution of the measurement close to $\Gamma_{\text{F}}$. On the right-hand side of Fig.~\ref{fig:charact}(b), we see the same measurement, now with the laser polarized linearly along the direction of $\bm{B}$. In this case, the transitions $E_2$ and $E_3$ almost completely vanish and we can estimate $s_{1}^{(1)} = s_{4}^{(1)} = 0.86(10)$ and $s_{2}^{(1)} = s_{3}^{(1)} = -0.86(10)$, with $s^{(1)}$ the rectilinear parameter of the Stokes vector (see Supplementary Material for details). This means that to a good approximation $s_1^{(1)},s_4^{(1)} \parallel \bm{B}$ and $s_2^{(1)},\bm{s}_3^{(1)} \perp \bm{B}$, as expected from Eq. \eqref{eq:relAngles}.

In the second step, we determine the electron-spin orientation of each transition directly using DNP \cite{Latta2009,Hogele2012}. To explain the dynamics of this process, we assume a (predominantly) isotropic $g^{\text{e}}_{\perp}$ and switch into the basis of the spin quantization axis $\hat{x}$ with a magnetic field $\bm{B} = B_x\,\hat{x}$. The Hamiltonian of the electron interacting with $N$ nuclei via the Fermi-contact hyperfine interaction is given by~\cite{Urbaszek2013}
\begin{equation}
    \mathcal{H}_{\text{e,n}} = g^{\text{e}}_{\perp} \mu_B B_x \hat{S}_x - \gamma_{\text{n}} B_x \sum_j^N \hat{I}_x^j + a \sum_j^N \hat{S}_x \hat{I}_x^j,
\label{eq:Hen}
\end{equation}
using the basis $\ket{S_x,I_x}$, with $\hat{S}_x$ the electron spin projection, $\hat{I}_x^j$ the spin-projection of the $j$-th nucleus, $\gamma_{\text{n}}$ the nuclear gyromagnetic ratio and $a$ the average hyperfine constant \cite{Urbaszek2013}. The energy spacing between two states with an average $I_x=\braket{\hat{I}_x}$ differing by a single nuclear-spin flip $\delta I_x = \pm 1$ is then
\begin{equation}
    E(S_x,I_x\pm 1) - E(S_x,I_x) = \pm\left(-\gamma_{\text{n}} B_x + a S_x\right)
\label{eq:DeltaE}.
\end{equation}
From Eq.~\eqref{eq:DeltaE}, we see that for the present case of $|\gamma_{\text{n}}|\ll |g^e_\perp \mu_B|$ and a nuclear Zeeman splitting $\hbar\,\omega^n_Z=|\gamma_{\text{n}} B_x| > a$, with $a>0$, the states of the manifold $\ket{\downarrow_x,I_x}$ ($\ket{\uparrow_x,I_x}$) exhibit an energy splitting of $\omega_Z^n$ increased (decreased) by $a$, as depicted in Fig.~\ref{fig:charact}(c). And, crucially, this difference of energy splittings only depends on the electron spin state $S_x$, not on the sign of the electron $g$-factor \cite{Puebla2013}. The trion is, to a good approximation, unaffected by the hyperfine interaction due to the electrons being in a singlet state and the hole's $p$-type Bloch function \cite{Fischer2008,Eble2009}. For a fixed change in nuclear polarization, the transition energy between the ground state electron and the trion thus increases or decreases depending on the electron spin state. Whether the optical transition energy increases or decreases with nuclear polarization thus acts as an absolute reference for defining the signs and relative magnitudes of $g^{\text{e}}_{\perp}$ and $g^{\text{t}}_{\perp}$. We measure this effect via optical resonance dragging \cite{Hogele2012}, as follows.

In the RF scheme shown in Fig.~\ref{fig:charact}(c) the laser is in resonance with the electron-trion transition $\hbar\omega_0 = \hbar\omega^\prime + a I_x^0$, with $\hbar\omega^\prime$ the electron-trion transition energy without hyperfine interaction and $I_x^0$ the steady-state nuclear polarization.
A small anisotropy in $g^{e}_{\perp}$ weakly allows sideband transitions with $\delta I_x = \pm 1$ to change the nuclear polarization \cite{shofer2024ArXiv}. When $S_x=+1/2$ and the optical transition energy decreases with nuclear polarization (upper panel), the side-band transitions involving a spin-flip towards $I_x^0$ are closer to resonance than sideband transitions involving a spin-flip away from $I_x^0$. This leads to an inward flow, towards a stable locking-point $I_x^0$. When $\omega_L$ changes slowly, the locking-point $\omega_0$ is ``dragged'' by building up nuclear polarization such that the resonance condition $\omega_0=\omega_L$ and thus the photon scattering level remains fixed \cite{Latta2009,Hogele2012}. At some point, the nuclear spin relaxation outpaces the weak side-band transitions, and the dragging abruptly ends. The opposite behavior is observed for the scenario shown in the lower part of Fig.~\ref{fig:charact}(c), where $S_x=-1/2$ and the optical transition energy increases with nuclear polarization. Here, the side-band transitions involving a spin-flip towards $I_x^0$ are farther from resonance than sideband transitions involving a spin-flip away from $I_x^0$, which favors a flow away from $I_x^0$. This leads to an "anti-dragging" behavior, where the nuclear polarization builds up in a way that avoids resonant scattering levels.

\begin{figure}
    \centering
    \includegraphics[scale = 1.25]{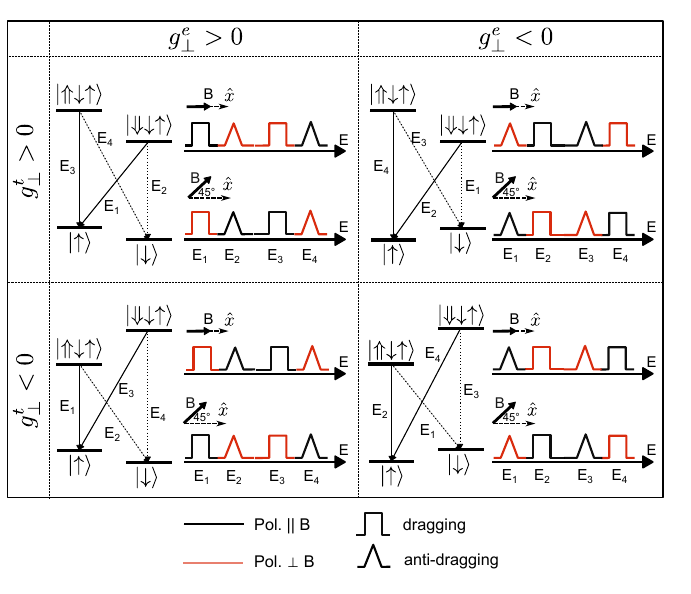}
    \caption{Table of the four possible configurations of transition polarization and dragging patterns for $|g^{\text{t}}|>|g^{\text{e}}|$ in the Voigt configuration, which can be used to identify the signs of the $g$-factor elements. The spin configurations are labeling the energy levels in the basis along the magnetic field $\bm{B}$ to provide an intuition for the $g$-factor signs.} 
    \label{fig:TruthTable}
\end{figure}

Figure~\ref{fig:charact}(d) shows the measurement of these dragged and anti-dragged transitions. The laser frequency $\omega_L/2\pi$ is slowly varied under a constant $V_G=\SI{0.93}{V}$, which is in the center of the first co-tunneling region seen in Fig.~\ref{fig:charact}(b). The resulting peaks marked with ``D'' are the dragged transitions associated with $S_x = +1/2$, showing the characteristic flat-top shape as the nuclear polarization builds up to follow the changing $\omega_0$. The peaks marked with ``A'' are the anti-dragged transitions associated with $S_x = -1/2$, identified by their more triangular (resonance-avoiding) shape and their often hysteretic behavior in the forward and backward frequency-scanning directions. This measurement unambiguously identifies that $E_1$ and $E_3$ ($E_2$ and $E_4$) are the transitions involving $\ket{S_x}=\ket{\uparrow}$ ($\ket{\downarrow}$).

With the two measurements above presented in Fig.~\ref{fig:charact}(b,d), we have all the information we need to fully identify $g^{\text{e,t}}_{\perp}$ of the QD. From Fig. \ref{fig:charact}(b), we concluded that $s_{1,4}^{(1)}\parallel \bm{B}$, which implies that $\mathcal{S}(g^{\text{e}}_{\perp})=\mathcal{S}(g^{\text{t}}_{\perp})$, according to Eq. \eqref{eq:relAngles}. The ordering of D and A transitions in Fig. \ref{fig:charact}(d) requires that that $|g^{\text{e}}_{\perp}|<|g^{\text{t}}_{\perp}|$ and pins down the $g$-factor signs to $g^{\text{e}}_{\perp}>0$ and $g^{\text{t}}_{\perp}>0$. Together with the measured energy splittings from Fig.~\ref{fig:charact}(b), we can determine $g^{\text{e}}_{\perp}=0.08(2)$ and $g^{\text{t}}_{\perp}=0.13(2)$. The electron $g$-factor is in excellent agreement with simulated $\braket{g^e_\perp}=0.09$ from Fig.~\ref{fig:simulation}(d) at an $X^-$ emission wavelength of \SI{772}{nm}. The hole $g$-factor agrees with the sign and is within a factor of 2 of the simulated value of$\braket{g^t_\perp}=0.27$. Strikingly, the alignment of the TDMs follows the expectation from Eq.~\ref{eq:relAngles} for ideal $D_{2d}$ symmetry.

\begin{figure*}
    \centering
    \includegraphics[scale = 1.03]{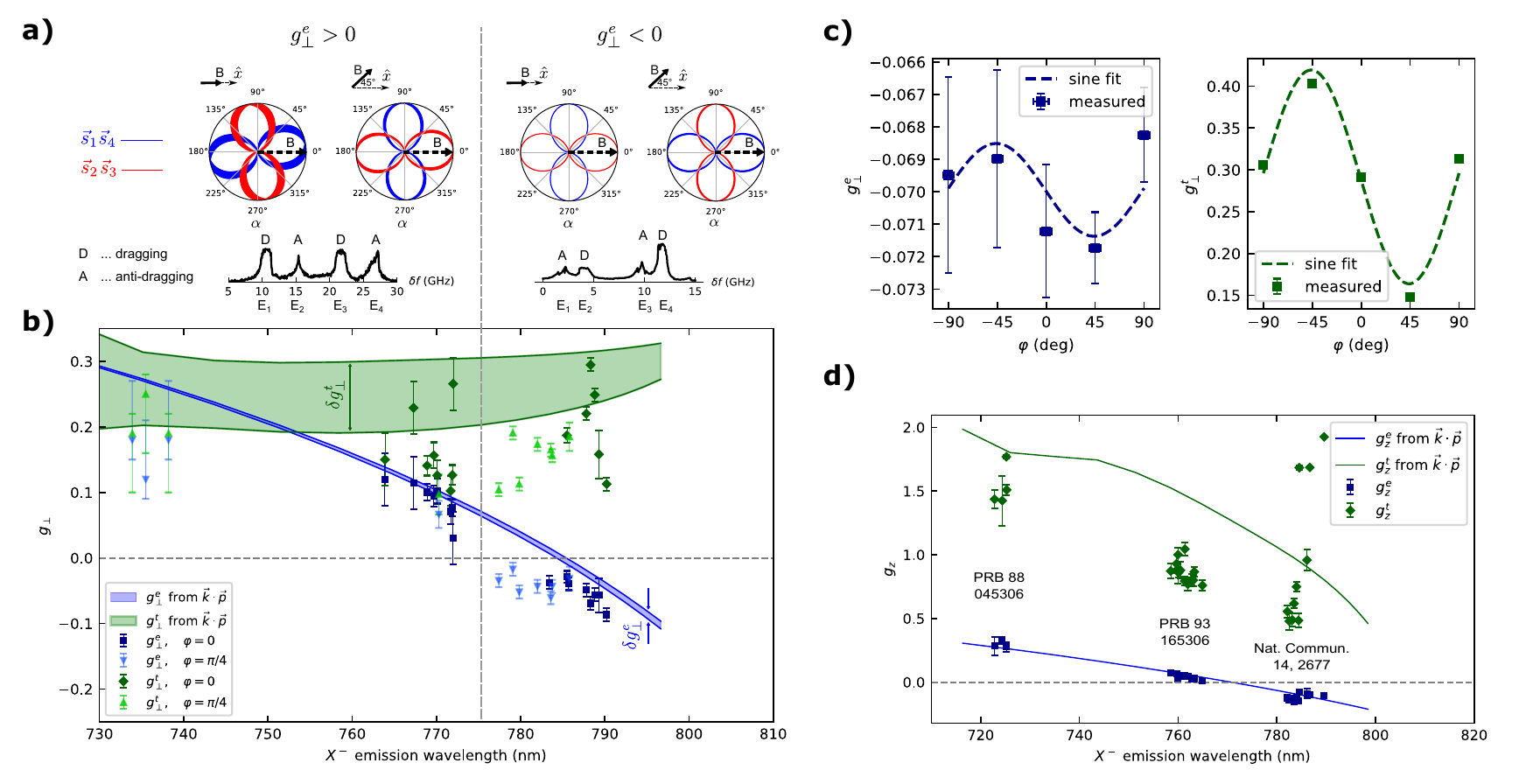}
    \caption{(a) Orientation of the optical transition dipole moments and dragging patterns for different $\varphi$ and $X^-$ emission wavelengths. The polar plots show an average of all measured $\bm{s}_{1,4}$ (blue) and $\bm{s}_{2,3}$ (red) oriented relative to $\bm{B}$ ($0^\circ$ reference). The dragging patterns below are measured at emission wavelengths of \SI{770}{nm} (left) and \SI{785}{nm} (right). (b) $g$-factors $g^{\text{e}}_{\perp}$ and $g^{\text{t}}_{\perp}$ of different LDE GaAs QDs across different samples measured in this work as a function of $X^-$ emission wavelength. The shaded areas show the range of values simulated via $\bm{k}\cdot\bm{p}$ across all in-plane directions, with the anisotropies indicated by $\delta g^e_\perp$ and $\delta g^t_\perp$. (c) In-plane $g$-factors of one QD emitting at $\SI{785}{nm}$ measured for different $\varphi$. (d) Compilation of Faraday $g$-factors $g^{\text{e}}_z$ and $g^{\text{t}}_z$ from literature \cite{Puebla2013,Ulhaq2016,Millington2023} together with our numerical predictions.} 
    \label{fig:summary}
\end{figure*}

The same measurement and analysis strategy can be applied to any QD system. Figure~\ref{fig:TruthTable} summarizes our measurement protocol, showing all four possible configurations of $g^{\text{e}}_{\perp}$ and $g^{\text{t}}_{\perp}$ which can be deduced from the measurements above (for $|g^{\text{e}}_{\perp}| < |g^{\text{t}}_{\perp}|$; the patterns for $|g^{\text{e}}_{\perp}| > |g^{\text{t}}_{\perp}|$ follow straight-forwardly).

\section{Applying the protocol to a broad range of QDs}
\label{sec:summary}

To verify the predictive capabilities of our simulations and to assess the reproducibility of LDE GaAs QDs, we characterize the in-plane $g$-factors $g^{e,t}_\perp$ and the TDMs of multiple LDE GaAs QDs across a wide range of transition wavelengths and samples coming from different growth facilities (see Supplementary Material for sample details) using our measurement protocol (section \ref{sec:measurement}). Figure \ref{fig:summary}(a) shows the mean values and standard deviations of the TDM orientation (indicated by the blurring around the polar axis) and the measured dragging patterns for different $\varphi$ and QD emission wavelengths. We observe a clear flip in both TDM orientation and dragging behavior between QDs with emission wavelength lower and higher than \SI{775}{nm}, which indicates the zero-crossing of $g_{\perp}^{e}$. The TDMs follow Eq.~\eqref{eq:relAngles} and our simulations closely -- the principal polarization axes are determined by the magnetic field orientation relative to the GaAs crystallographic axes -- which confirms that in all measured LDE GaAs QDs the $D_{2d}$-symmetric behavior dominates the orientation of the TDMs. 

We can estimate the robustness of Eq.~\eqref{eq:relAngles} against geometric imperfections by taking the fine structure splitting (FSS) of the neutral exciton as a measure of the in-plane anisotropy since a high FSS is an indicator for an ellipticity of the QD shape in the $\hat{x}$-$\hat{y}$ plane \cite{Plumhof2010}. The highest FSS value measured across all QDs in this work is \SI{3}{GHz} (\SI{12}{\micro eV}), which is an outlier among modern LDE GaAs QDs with typical values of $\leq\SI{4}{\micro eV}$ \cite{Huo2013FSS,Schimpf2021,BassoBasset2021}. In this outlier case, we see a deviation of the TDM orientation of about \SI{20}{\degree} away from $\bm{B}$, while Eq.~\eqref{eq:relAngles} predicts the TDM to be aligned with $\bm{B}$ (see Supplementary Material for details). For the other QDs in this study with FSS values $\lesssim\SI{1.2}{GHz}$ (\SI{5}{\micro eV}) we observe deviations of $\lesssim\SI{10}{\degree}$. This relatively weak dependence on the FSS suggests the robustness of the TDMs against normal variations in QD shape and that the neutral exciton FSS can be used as a proxy for the expected correction to the ideal TDM angle.

Figure \ref{fig:summary}(b) summarizes the values of $g_{\perp}^{e}$ and $g_{\perp}^{t}$ as a function of $X^-$ emission wavelength, with the correct signs deduced from the measured TDMs and dragging patterns. We note that the dragging patterns were measured for a subset of QDs with $g_{\perp}^{e}$ around zero, where the ambiguity of the $g$-factor is the highest. For QDs with wavelengths far away from the zero-crossing, it is reasonable to assume that the sign of $g_{\perp}^{e}$ is the same as that determined close to and on that side of the zero-crossing. Remarkably, most measured values of $g_{\perp}^{e}$ agree within confidence bounds with the trend predicted by our simulations (shaded blue region). For the trions, $g_{\perp}^{t}$ follows the prediction from the simulation of a positive value that is approximately independent of wavelength. While this value is larger by $\sim 50\%$ in our simulation compared to our measurement of $\braket{g_{\perp}^{t}}\approx 0.18$, this remains a useful predictor given the complex nature of the in-plane trion $g$-factor and the range of values it could have conceivably acquired (see Sec.\,\ref{sec:theory}). The green shaded region depicts the range of simulated values obtained as a function of magnetic field angle relative to crystallographic axes and thus represents a significant anisotropy of $g_{\perp}^{t}$. Our measurements do not distinguish between the [110] and the [1$\bar{1}$0] (which is possible in principle, by observing patterns on the back of the unpolished wafer substrates or by keeping track of the crystal orientation during wafer processing), which may contribute to the scatter of our experimental data. 

To get a better measurement of both electron and trion anisotropies we picked one QD with an emission wavelength of $\SI{785}{nm}$ and measured $g^{\text{e}}_{\perp}$ and $g^{\text{t}}_{\perp}$ for different $\varphi$, as shown in Fig.~\ref{fig:summary}(c). We observe that for the electron (left) the anisotropy $\delta g^{\text{e}}_{\perp}=0.003(2)$ is barely resolvable in our measurements, but non-negligible, as expected from our simulation in Sec. \ref{sec:sim} and from previous studies \cite{shofer2024ArXiv}. The trion (right) shows a more pronounced anisotropy of $\delta g^{\text{t}}_{\perp} = 0.25(3)$, with the minima and maxima clearly aligned with the [110] ($\phi=\SI{45}{\degree}$) and [1$\overline{1}$0] ($\phi=\SI{135}{\degree}$) axes, as expected from the symmetry of the zinc-blende crystal structure \cite{Luo2015} and from our simulations. While the measured value of $\delta g^{\text{t}} \sim 0.25$ is higher than the 0.15 expected from our simulations, the prediction of the magnetic properties of the trion within less than a factor of 2 remains a key result.

To complete the picture, we show a compilation of $g^{\text{e,t}}_z$ values from different LDE GaAs QDs in the literature \cite{Puebla2013,Ulhaq2016,Millington2023} in Figure \ref{fig:summary}(d). These values were measured via the magnetic field-dependent splittings of the neutral exciton emission lines (see Ref.~\onlinecite{Puebla2013} for details) and also follow our simulations closely.

\section{Conclusion}

Our numerical and experimental studies on GaAs QDs obtained by local droplet etching epitaxy (LDE GaAs QDs) show that, for the electrons with spherically symmetric $s$-like Bloch functions, the $g$-tensor can be well described by the balance between the intrinsic magnetic moment and spin-correlated orbital currents (SCOCs), which are defined by the wavefunction envelope \cite{VanBree2014}. We predict numerically and verify experimentally that the electron $g$-factor crosses zero at an emission wavelength of about \SI{770}{nm} in the Faraday configuration (out-of-plane magnetic field) and about \SI{780}{nm} in the Voigt configuration (in-plane magnetic field). Our polarization- and electron-spin-sensitive resonance fluorescence measurements over wavelengths from \SI{730}{nm} to \SI{790}{nm} remove any remaining ambiguity about $g$-factor signs, even at values very close to zero. The measured electron- and trion- $g$-factors and the associated optical transition dipole moments (TDMs) follow our predictions from the multiband $\bm{k}\cdot\bm{p}$-configuration-interaction framework, which only uses the AFM-measured shape of one Al droplet-etched nanohole as input.

The more complex behavior of trions emerges from the $p$-like Bloch function of the heavy hole (HH) and the distinction of the growth axis in zinc-blende heterostructures. In the Voigt configuration, the Zeeman interaction vanishes for the HH, and the magnetic response and the relative spin phase of the trion wavefunction are dominated by higher-order corrections -- likely a combination of a non-Zeeman term and LH admixture (see Supplementary Material). This leads to a non-trivial dependence of the orientation of the TDMs on the in-plane component of the external magnetic field \cite{Pikus1994,Semenov2003,Trifonov2021}. However, high geometrical symmetry, QD material homogeneity, and low built-in strain in LDE GaAs QDs \cite{DaSilva2021,Schimpf2021,BassoBasset2021} make the TDMs predictable and tunable with magnetic field by assuming $D_{2d}$ symmetry \cite{Semenov2003,Trifonov2021}, despite the intersection of the underlying zinc blende crystal symmetry and the conical shape symmetry being formally a lower $C_{2v}$ symmetry. With the fine structure splitting (FSS) indicative of the degree of in-plane asymmetry \cite{Huo2013FSS}, we can estimate that the $D_{2d}$ approximation holds for QDs with an FSS at least up to about \SI{2.5}{GHz} (\SI{10}{\micro eV}), well above typical for those QDs. The reduced $C_{2v}$ symmetry is revealed mainly as a broken in-plane symmetry in the vicinity of the GaAs-AlGaAs interface \cite{Luo2015}, which leads to the observed anisotropies of the in-plane $g$-factors \cite{Kusrayev1999} and to heavy hole-light hole mixing (HH-LH mixing), even in the absence of strain or geometric imperfections. We find that in unstrained LDE GaAs QDs, this HH-LH mixing is mostly confined to the region close to the GaAs/AlGaAs interface, where the electron-hole wavefunction overlap is small, and thus the impact of HH-LH mixing on the optical properties is almost negligible as evidenced by 99.98\% degree of circular polarization of the trion emission without the magnetic field. 

In conclusion, we have demonstrated that the spin-photon interface in LDE GaAs QDs can be predicted and designed before QD growth and then realized by choosing the appropriate GaAs filling height and Al concentration in the barrier. This is made possible by the strain-free and reproducible growth mode, which yields highly in-plane symmetric and homogeneous QDs with a spread of emission wavelengths $<\SI{5}{nm}$ across the wafer \cite{DaSilva2021}. The $g$-tensors governing the relevant energy splittings and the TDMs can be measured without ambiguity, even for values close to zero, following the protocol introduced in this work. The TDMs have a well-defined dependency on the external magnetic field vector, which can be exploited to tune the polarization and the cyclicity of the optical transitions and the electron-nuclear interface during operation. We believe that our findings will vastly facilitate the design of high-quality spin-qubits in LDE GaAs QDs for highly integrated and scalable quantum technology.

\begin{acknowledgements}
This research was funded by the Austrian Science Fund (FWF) 10.55776/J4784 (C.S.) and via the Research Group FG5 (A.R.) and the cluster of excellence quantA [10.55776/COE1; A.R.]. We acknowledge support from ETRI through The Institute of Information \& communications Technology Planning \& Evaluation (IITP) grant funded by the Korea government (MSIT) (No.2022-0-00463, Development of a quantum repeater in optical fiber networks for quantum internet (C.S., D.A.G.), the EPSRC New Investigator Award EP/W035839/2 (D.A.G., Z.X.K.), and the QuantERA project MEEDGARD through EPSRC EP/Z000556/1 (D.A.G., M.A., Y.K.), the European Union’s Horizon 2020 research and innovation program under Grant Agreement No. 871130 (Ascent+; A.R.) and the EU HE EIC Pathfinder challenges action under grant agreement No. 101115575 (Q-ONE; A.R.), as well as the Linz Institute of Technology Secure and Correct Systems Lab, supported by the State of Upper Austria (A.R.) and the FFG (Grants No. 891366 and 906046; A.R.). A.H. acknowledges an EPSRC DTP studentship EP/W524311/1. D.A.G acknowledges a Royal Society University Research Fellowship. M.~G. acknowledges the financing of the MEEDGARD project funded within the QuantERA II Program that has received funding from the European Union's Horizon 2020 research and innovation program under Grant Agreement No. 101017733 and National Centre for Research and Development, Poland --- project No. QUANTERAII/2/56/MEEDGARD/2024. Part of the calculations have been carried out using resources provided by the Wroc{\l}aw Centre for Networking and Supercomputing, Grant No. 203 (M.G.). The work in Basel was funded by Swiss National Science Foundation via project sQnet (20QU-1_215955). M.M., A.L. and P.S. thank the French RENATECH network. S.c.d.S. acknowledges funding by the São Paulo Research Foundation (FAPESP) under the grant numbers 2024/08527-2 and 2024/21615-8.\par
C.S., M.H.A., Z.X.K., and D.A.G. thank Evgeny A. Chekhovich for helping to compile previous measurements from literature and for many fruitful discussions. M.G. is grateful to K. Gawarecki for sharing his computational code. G.N.N., L.L.N. and R.J.W. thank Clemens Spinnler and Marcel Erbe for expert assistance in the experiments carried out in Basel. \par
\end{acknowledgements}


\bibliography{main}

\end{document}


\preprint{APS/123-QED}

\title{Optical and magnetic response by design in GaAs quantum dots\\
        Supplementary material}

\maketitle
\tableofcontents

\newpage
\section{Theoretical considerations}

\subsection{Electron-trion transition dipole moments in a magnetic field}
\label{sec:TDM}

For our qualitative considerations, we treat the electron states as well-separated from other bands in the semiconductor, and we describe them in the basis of spinors $\ket{\psi^\text{e}_1}=\ket{\uparrow}\otimes\ket{S}$ and $\ket{\psi^\text{e}_2}=\ket{\downarrow}\otimes\ket{S}$, where $S$ denotes the spherically symmetric Bloch function at the $\Gamma$ point. As we are mainly interested in the magnetic and optical response, we approximate the trion by a single valence-band hole as the two electrons in the trion form a spin-singlet. We neglect the spin-orbit split-off subband. The light-hole (LH) subband is split from the heavy-hole (HH) by the confinement and strain distinguishing the growth axis in the crystal. This allows us to treat the predominantly HH hole states as a pseudo-spin-1/2 system with basis states $\ket{\psi^\text{h}_1}= \ket{\uparrow}\otimes \left(\ket{X} - i\ket{Y} \right)$ and $\ket{\psi^\text{h}_2}=- \ket{\downarrow}\otimes \left(\ket{X} + i\ket{Y} \right)$, where $X$ and $Y$ are again the Bloch functions \cite{Trifonov2021,Chuang2009}. 
Note that even though we discuss holes, we work here in the \textit{electron} picture, i.e., the states and Hamiltonians we write describe the electron states in the valence band with which the hole is associated. Further, in this manner, we also calculate optical transitions as transitions of an electron between bound states formed in conduction and valence bands. Switching to the \textit{hole} picture could be done by applying time reversal to hole/valence Hamiltonians and states and looking for the optical transition between an electron-hole pair and vacuum. The impact of a magnetic field on electron states is governed by the regular Zeeman Hamiltonian,
\begin{equation}
    \mathcal{H}^{\text{e}}(\bm{B}) = \frac{1}{2} \mu_{\text{B}} \left[g^{\text{e}}_z B_z \sigma_z + g^{\text{e}}_{\perp} \left( B_x\sigma_x + B_y\sigma_y \right)\right],
\label{eq:HZe}
\end{equation}
in the coordinate frame $\{x,y,z\} := \{[100],[010],[001]\}$, with $\mu_{\text{B}}$ the Bohr magneton, $\sigma_i$ the Pauli matrices, $g^{\text{e}}_z$ the out-of-plane and $g^{\text{e}}_{\perp}$ the in-plane electron $g$-factors, where here for simplicity we assume in-plane isotropy.

The HH, and thus trion, response to the magnetic fields needs more attention. The direct impact of the field on the HH and LH valence-band states is described by the Hamiltonian
\begin{equation}
H_{\text{Z}} = -2\mu_{\text{B}}\left( \kappa \, \bm{J}\cdot\bm{B}+q \, \bm{\mathcal{J}}\cdot\bm{B}\right),
\label{eq:HZval}
\end{equation}
with $\bm{J}=(J_x,J_y,J_z)$, $\bm{\mathcal{J}}=(J_x^2,J_y^3,J_z^3)$, $J_i$ the angular momentum 3/2 matrices and $\kappa$ and $q$ the constants weighting the contributions. The Hamiltonian contains the standard isotropic Zeeman term and the anisotropic term originating from the crystal tetrahedral symmetry (point group $T_d$) lower than spherical. Assuming no further reduction of symmetry, the Zeeman term produces a splitting in the Faraday configuration but does not contribute in the first order to the response in the Voigt configuration. However, the anisotropic term does, and together they lead to a Zeeman-like (but of opposite phase) behavior discussed further. This situation also holds for the maximal possible symmetry of a QD formed in the zinc-blende semiconductor, which is described by the point group $D_{2d}$ that arises due to the distinction of the growth [001] axis by the heterostructure. At lower symmetries, other contributions to the hole/trion response arise. Nonetheless, using the simplified basis, we are still able to generally write the trion eigenstates formally in the same form as the electron Zeeman eigenstates
\begin{equation}
\begin{aligned}
    \ket{\psi_+^j(\chi,\theta)} &{}= \alpha^j(\chi)\,e^{-i\frac{\theta}{2}}\,\ket{\psi^j_1} + \beta^j(\chi)\, e^{i\frac{\theta}{2}}\, \ket{\psi^j_2},\\
    \ket{\psi_-^j(\chi,\theta)} &{}= -\beta^j(\chi)\,e^{-i\frac{\theta}{2}}\,\ket{\psi^j_1} + \alpha^j(\chi)\, e^{i\frac{\theta}{2}}\, \ket{\psi^j_2},
\end{aligned}
\end{equation}
with $j \in \{\text{e},\text{h}\}$, $\theta$ a generic phase, $\chi$ as defined in the main text, and the coefficients
\begin{equation}
\begin{aligned}
   \alpha^j &= \frac{1}{\mathcal{N}}\left(g^j_z \cos{\chi} + c\right),\\
   \beta^j &= \frac{1}{\mathcal{N}} \left(g^j_{\perp} \sin{\chi}\right),\\
   \mathcal{N} &= \sqrt{2 \left( c^2 + c\, g^j_z \cos{\chi}  \right)},\\
   c &=  \sqrt{(g^j_z \cos{\chi})^2 + (g^j_{\perp} \sin{\chi})^2}.
\end{aligned}
\end{equation}
For isotropic $g_\perp^e$ the in-plane component of the quantization axis of the $s$-like spin 1/2 electron lies directly along $\bm{B}$, so that $\theta=\varphi$ for the electron states ($\varphi$ as defined in the main text). For the trion, we keep the generic phase $\theta$ to be able to accommodate general solutions due to perturbations that we introduce further.

We can calculate the optical transition dipole moment matrix elements at an in-plane angle of the electric field (linear polarization) $\alpha$ relative to $\bm{B}$ (i.e., $\alpha+\varphi$ relative to $\hat{x}$) via the dipole moment operator \cite{Kusrayev1999,Semenov2003}
\begin{equation}
\hat{V}_{\varphi,\alpha} = \frac{1}{2} \left[\left(\hat{p}_x + i\hat{p}_y\right) e^{i(\varphi+\alpha)} + \left(\hat{p}_x - i\hat{p}_y\right) e^{-i(\varphi+\alpha)}\right],
\end{equation}
where $\bra{S}\hat{p}_x \ket{X} = \bra{S}\hat{p}_y\ket{Y} = m_0 P / \hbar$ ($P$ is the interband momentum matrix element) and $\bra{S}\hat{p}_x\ket{Y} = \bra{S}\hat{p}_y \ket{X} = 0$ due to Bloch functions' symmetry. The rates of the four possible transitions are then given by $p^{i,k}_{\varphi,\alpha,\theta}\propto|M^{i,k}_{\varphi,\alpha}|^2$, with $M^{i,k}_{\varphi,\alpha}=\braket{\psi_{i}^{\text{e}} |\hat{V}_{\varphi,\alpha}| \psi_{k}^{\text{h}}}$ and $i,k \in \{+,-\}$. Note that $\hat{V}_{\varphi,\alpha}$ acts on both entries of the spinor independently. In the Voigt configuration, $\chi=\pi/2$, all four transitions are linearly polarized and equally allowed, with the polarization behaving as
\begin{equation}
    p^{i,k}_{\varphi,\alpha} \propto \begin{cases}
  \text{sin}^2[3\varphi/2 + \alpha -\theta(\varphi)/2] & \text{for }i=k\\    
  \text{cos}^2[3\varphi/2 + \alpha -\theta(\varphi)/2]  & \text{for }i\neq k,
\end{cases}
\label{eq:pVoigt}
\end{equation}
where $\theta(\varphi)$ is the not yet determined relative phase of the HH state superposition, which is some function of $\varphi$. The Stokes vectors along the optical axis [001] and with the linear components aligned to $\bm{B}_\perp$ (called $\bm{s}_{1-4}$ in the main text) can then readily be calculated as:
\begin{equation}
\bm{S}=
\begin{pmatrix}
    S^{(0)}\\S^{(1)}\\S^{(2)}\\S^{(3)}
\end{pmatrix}=
\begin{pmatrix}
    p^{i,k}_{\varphi,0} + p^{i,k}_{\varphi,\pi/2}\\
    p^{i,k}_{\varphi,0} - p^{i,k}_{\varphi,\pi/2}\\
    2\,\text{Re}(M_{\varphi,0} M_{\varphi,\pi/2}^\ast)\\
    -2\,\text{Im}(M_{\varphi,0} M_{\varphi,\pi/2}^\ast)
\end{pmatrix}.
\end{equation}

The $p$-like Boch functions of the HHs align with the (001) plane so that first-order Zeeman terms are prohibited in Voigt geometry. To find $\theta$ and to understand possible anisotropies of $g^t_{\perp}$, we briefly discuss the three possible interactions between the HH and a magnetic field, their dependencies, and their impact on the optical transition dipole moments (more details can be found in Ref.~\cite{Semenov2003} and references therein):

\begin{itemize}
    \item The third-order Zeeman interaction (third-order perturbation within the Zeeman term in Eq.~\eqref{eq:HZval}) yields an effective term
    \begin{equation}
        V^{(3)}_\text{Z} = \frac{3}{2}\frac{(\mu_{\text{B}}\kappa B)^3}{\Delta_{\text{LH}}^2}\left[ \sigma_x \cos(3\varphi) + \sigma_y\sin(3\varphi) \right]
    \end{equation}
    where
    $\Delta_{\text{LH}}$ is the HH-LH splitting. This term yields an isotropic contribution to $g^t_\perp$ and $\theta=3\varphi$. Considering Eq.~\eqref{eq:pVoigt}, this leads to a linear polarization which follows $\bm{B}$ independent of $\varphi$. However, this term is very weak at reasonable fields, as it is cubic in the magnetic field.

    \item The Luttinger Hamiltonian allows for non-Zeeman magnetic interactions given by the second term of Eq.~\eqref{eq:HZval} 
    This Hamiltonian has a direct first-order matrix element between the HH states that can be used to write an effective Hamiltonian in the pseudo-spin HH basis
    \begin{equation}
         V^{(1)}_{q} = -\frac{3}{4} q \mu_{\text{B}}B (\sigma_x \cos{\varphi} - \sigma_y \sin{\varphi}),
    \end{equation}
    which yields an isotropic contribution to $g^t_\perp$ and $\theta=-\varphi$. The factor $q$ is small in bulk GaAs and in quantum wells, but it gets effectively strongly enhanced in QDs, as the translational symmetry is broken and $k$ is not a good quantum number anymore \cite{Golub1999,Trifonov2021}. Taking into account the negligible impact of the third-order term, this contribution is the main one present for holes, and thus trions, in the Voigt configuration in ideally symmetric disk-shaped quantum dots representing $D_{2d}$ symmetry. This contribution leads to the dominant polarization behavior observed and described in the main text.

    \item At the lower overall symmetry of the QD, the HH-LH mixing effect comes into play. Formally, the lack of [001]-axis inversion is enough to reduce the symmetry to $C_{2v}$. However, the mixing effects become important when there is an in-plane asymmetry of the confinement potential. The effective Hamiltonian for HH spins arises from a second-order perturbation in the combined Zeeman term from Eq.~\eqref{eq:HZval} and the HH-LH mixing term of the form $\propto t(J_xJ_y+J_yJ_x)$
    \begin{equation}
        V^{(2)}_{ht} = -\frac{3}{2} \mu_{\text{B}}B \, t \left( \sigma_x\,\sin\varphi -\sigma_y\,\cos\varphi \right),
    \end{equation}
    with $t$ a general coupling factor, which is inversely proportional to $\Delta_{\text{LH}}$. $V^{(2)}_{ht}$ leads to anisotropic contribution $\pm \delta g^{\text{t}}_{\perp}$ to $g^t_\perp$ that, in the absence of strain or other additional symmetry-braking elements, is aligned along the [110] and [1$\bar{1}$0] crystal axes. The fact that we see the anisotropy both in simulation and measurement consistently along these axes supports the argument that the in-plane uniaxial strain component in these dots is either negligible or consistently aligned with the [110] and [1$\bar{1}$0] axes. This term contributes to the wave-function phase as $\theta=\varphi+\pi/2$, which pins the two linear polarization components of the emission to the [110] and the [1$\bar{-1}$0] axes, entirely independent of the direction of $\bm{B}$.
\end{itemize}

\begin{figure*}[h]
    \centering
    \includegraphics[scale=1]{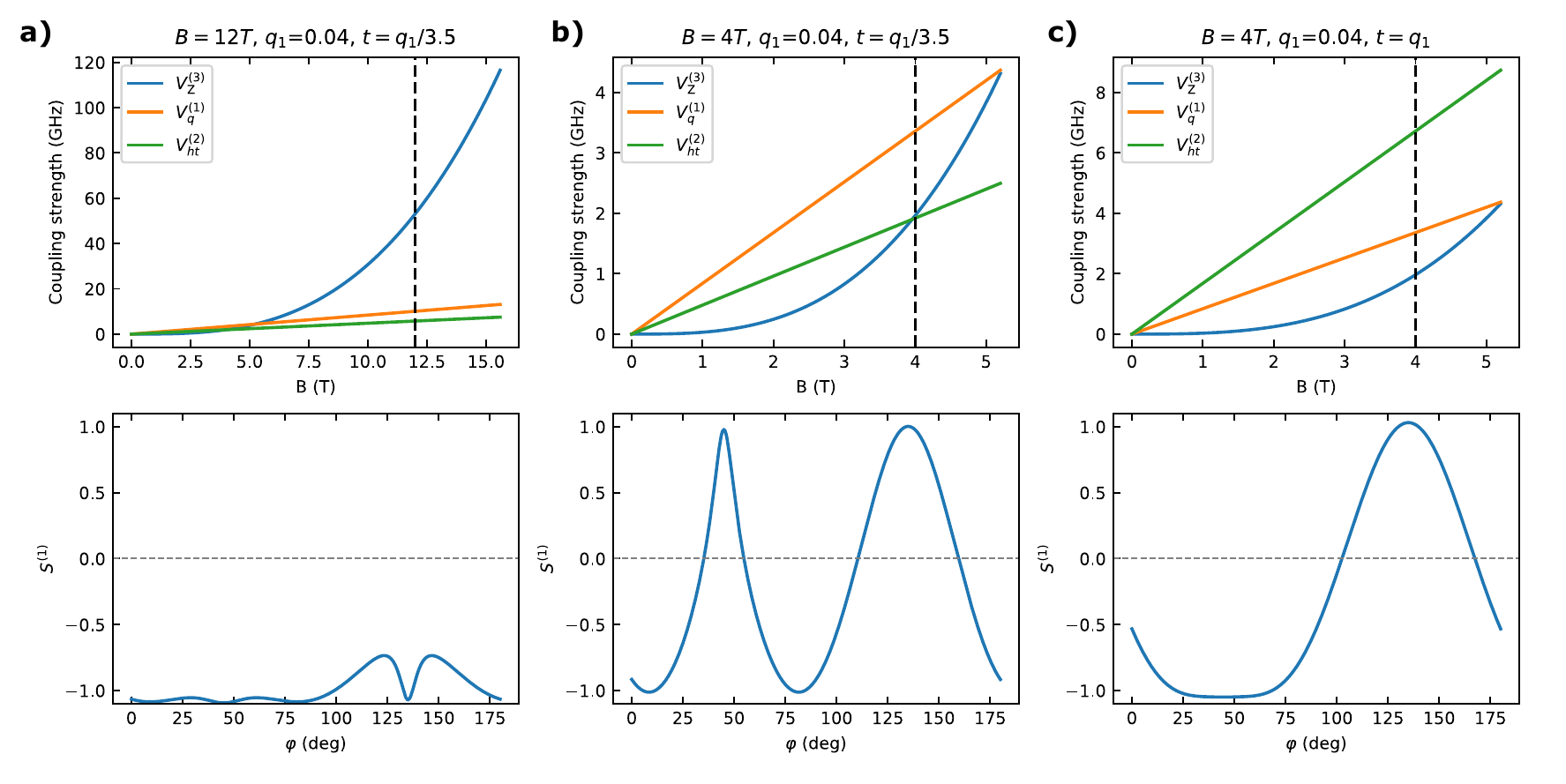}
    \caption{Different configurations of HH interaction contributions and their action on the transition dipole moments, using $\kappa=1.28$. (a) Dominant third order Zeeman term $V^{(3)}_\text{Z}$ at high magnetic fields. (b) Dominant non-Zeeman term $V^{(1)}_{q}$ as observed in this work. (c) Dominant $V^{(2)}_{ht}$, which pins the dipoles to the [110] and [1$\bar{1}$0] crystallographic directions. }
    \label{fig:HHHamiltonian}
\end{figure*}

To get an intuition for the weights of these three contributions and their dependency on $B$, $q$ and $t$, we take a qualitative look at the linear polarization $S^{(1)}$ for different $\varphi$, which we model by following Eq. (29) in Ref. \cite{Semenov2003}. The results for an assumed $q=0.03$ and three different parameter configurations are shown in Fig.~\ref{fig:HHHamiltonian}. In Fig.~\ref{fig:HHHamiltonian}(a), we assume a high magnetic field of \SI{12}{T}, for which the coupling strength from the third order Zeeman term $V^{(3)}_\text{Z}$ dominates due to the cubic scaling with $B$. As a consequence, the TDMs follow the magnetic field, and $S^{(1)}$ is largely independent of $\varphi$, i.e., the polarization follows the magnetic field vector. The case shown in Fig.~\ref{fig:HHHamiltonian}(b) assumes a magnetic field of \SI{4}{T}, which is commonly used in our experiments. Here, the non-Zeeman interaction  $V^{(1)}_{q}$ dominates, which leads to the dependency of $S^{(1)}$ on $\varphi$ as observed and described in the main text. In the hypothetical scenario of Fig.~\ref{fig:HHHamiltonian}(c), we set $t$ equal to $q$, which pins the TDMs to the [110] and [1$\bar{1}$0] crystallographic directions. 

\begin{figure*}[h]
    \centering
    \includegraphics[]{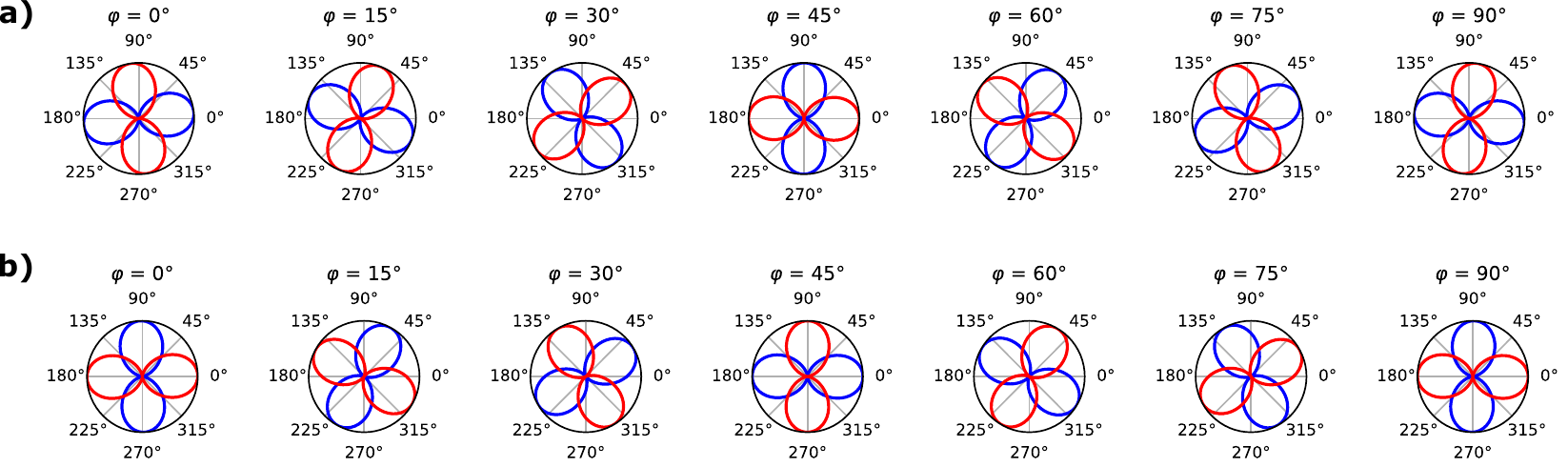}
    \caption{Linear polarizations of the dipole transition moments relative to $\bm{B}$ calculated from $\bm{k}\cdot\bm{p}$ for different $\varphi$ for (a) $g^e_\perp=0.04$, $g^t_\perp=0.17$ (b) $g^e_\perp=-0.08$, $g^t_\perp=0.16$.}
    \label{fig:simDipoles}
\end{figure*}

Under the assumption that $V^{(1)}_{q}$ dominates the trion spin superposition relative phase, we can set $\theta=-\varphi$, which results in the polarization 
\begin{equation}
    p^{i,k}_{\varphi,\alpha} \propto \begin{cases}
  \text{sin}^2(\varphi + \alpha) & \text{for }i=k,\\    
  \text{cos}^2(\varphi + \alpha) & \text{for }i\neq k,
\end{cases}
\label{eq:pVq}
\end{equation}
which describes the behavior described in the main text. Figure \ref{fig:simDipoles} depicts the TDM alignments calculated from $\bm{k}\cdot\bm{p}$ (more details in Sec. \ref{sec:simulations}) at angles relative to $\bm{B}$ for different $\varphi$. Both the behaviors in Fig. \ref{fig:simDipoles}(a) for $g^t_\perp>0$ and $g^e_\perp>0$, and in Fig. \ref{fig:simDipoles}(b) for $g^t_\perp>0$ and $g^e_\perp<0$ are consistent with the ones derived in this section under the assumption of $D_{2d}$ symmetry.

\section{Additional information to numerical simulations}
\label{sec:simulations}

\subsection{Material composition and strain distributions}

\begin{figure*}[h]
    \centering
    \includegraphics[]{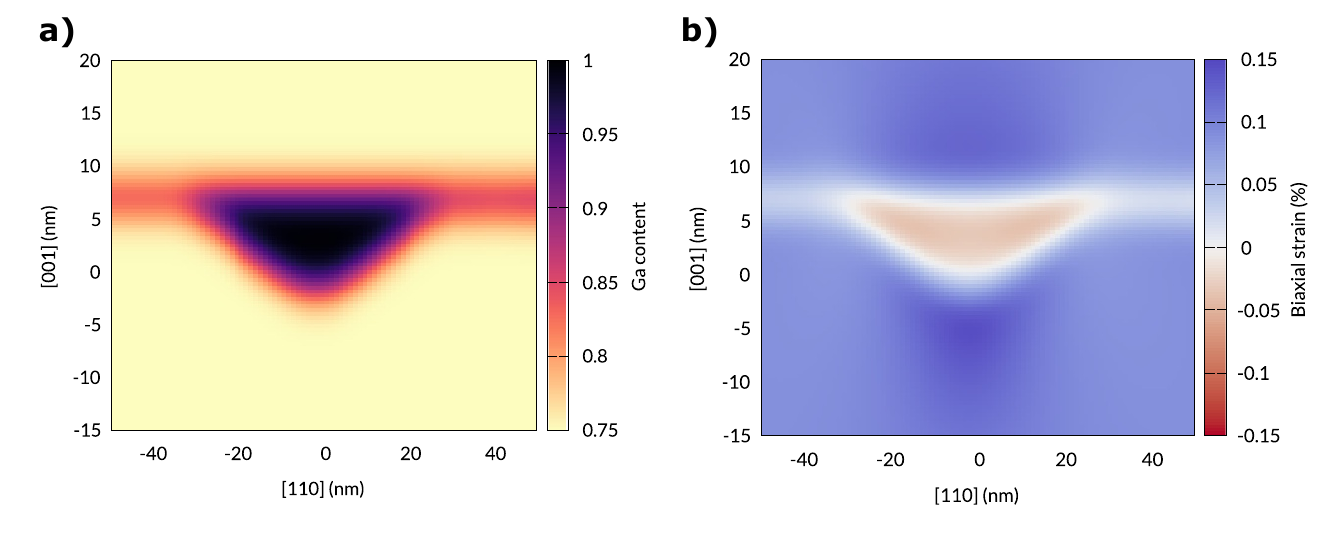}
    \caption{(a) Cross-section of the GaAs/AlGaAs material composition in a QD used in the $\bm{k}\cdot\bm{p}$ simulations. (b) Calculated biaxial strain distribution in the QD and in the barrier.}
    \label{fig:simDistribution}
\end{figure*}

Figure \ref{fig:simDistribution}(a) shows the spatial distribution of Ga concentration in the GaAs/AlGaAs heterostructure used in the $\bm{k}\cdot\bm{p}$ simulations. The QD shape is taken directly from the AFM nanohole scan. Realistic interfaces between the GaAs QD and the AlGaAs barrier are modeled by applying Gaussian averaging appropriate for normal diffusion. The spatial extent of this diffusion equal 1.5~nm is based on the known fraction of Al atoms that the electron wave function overlaps with in such a QD \cite{Zaporski2023}. The structure is grown on a GaAs substrate, which should lead to built-in strain in all material layers above it, which can be partially relaxed at subsequent interfaces. Based on the biaxial stain-induced quadrupolar shifts in nuclear magnetic resonance data \cite{Ulhaq2016}, we can estimate that half of this strain is present in the sample layer containing QDs. We apply this built-in strain as a boundary condition at the bottom wall of the computational box.

By minimizing the elastic energy within the continuum elasticity theory, we calculate the strain in the structure. Figure \ref{fig:simDistribution}(b) shows the distribution of bi-axial strain in the QD and the surrounding barrier material. The strain in the structure is very low due to the low lattice mismatch of materials used, and we focus on biaxial strain only, as it splits the heavy and light hole bands and is thus essential for HH-LH mixing effects. The AlGaAs barrier has a slightly larger lattice constant than the pure GaAs substrate below, which results in residual compressive strain in the barrier. The GaAs QD, embedded in the barrier, experiences tensile strain due to the lattice mismatch with the barrier. At the interfaces, a zone of zero strain emerges, which leads to particularly HH-LH-mixing rich areas close to the boundaries, which we will describe in more detail in the following.

\subsection{Electron and trion eigenstates}

For the simulated QD structure, we find the electron and hole eigenstates using a custom implementation \cite{Gawarecki2014} of the eight-band $\bm{k}\cdot\bm{p}$ method \cite{Bahder1992} that includes the strain, shear strain-induced piezoelectric field up to second order terms in the strain components, spin-orbit effects and arbitrary external electric, magnetic and strain fields. Significantly for this work focused on the magnetic response, the inclusion of the magnetic field is implemented using a gauge-invariant discretization scheme \cite{Andlauer2008}.

Finding electron $g$-factors is straightforward, as it only requires the calculation of the electron ground state Zeeman doublet in a magnetic field. The trion consists of two electrons forming a spin singlet and a hole, so the first approximation would be to calculate the trion $g$-factor just based on the hole ground state spin doublet. While electron spins do not contribute to the trion magnetic response, the presence of electrons does contribute via the many-particle interactions that lead to the hole in the trion being composed not only of its ground state but also having higher hole level admixtures, which renormalizes the trion magnetic response. For this reason, we use the configuration-interaction method to find the negative trion eigenstates. To this end, we form a configuration basis out of 20 electron and 26 hole single-particle eigenstates. While for converging the carrier-complex energy using a larger basis (even containing computational box states when there are no more states well confined in a QD) is always favorable, for quantities like the magnetic response, care is needed in this aspect. For this reason, we limit the basis to states well confined in a QD, as those higher states that start to leak to the computational box exhibit an unphysical magnetic response, which translates into an onset of unexpected oscillations of calculated $g$-factors versus the basis size. 

\subsection{Heavy-light hole mixing}

\begin{figure*}[h]
    \centering
    \includegraphics[]{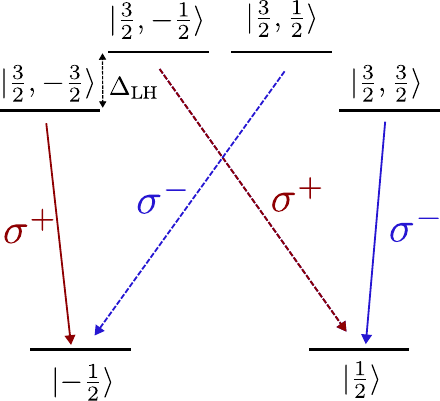}
    \caption{Electron-trion energy level scheme including HHs and LHs.}
    \label{fig:MixingOptics}
\end{figure*}

Figure \ref{fig:MixingOptics} schematically depicts the electron-trion energy levels with both the orbital ground states dominated by HHs $\ket{\frac{3}{2},\pm\frac{3}{2}}$ ($\ket{\Uparrow}$, $\ket{\Downarrow}$) and LHs $\ket{\frac{3}{2},\pm\frac{1}{2}}$ ($\ket{\uparrow}$, $\ket{\downarrow}$) present, separated by the HH-LH splitting $\Delta_{\text{LH}}$. In an actual system, those subbands gradually hybridize along the trion energy level ladder, which can be understood as a result of HH-LH mixing effects on the idealized levels of Fig.~\ref{fig:MixingOptics}.

\begin{figure*}[h]
    \centering
    \includegraphics[]{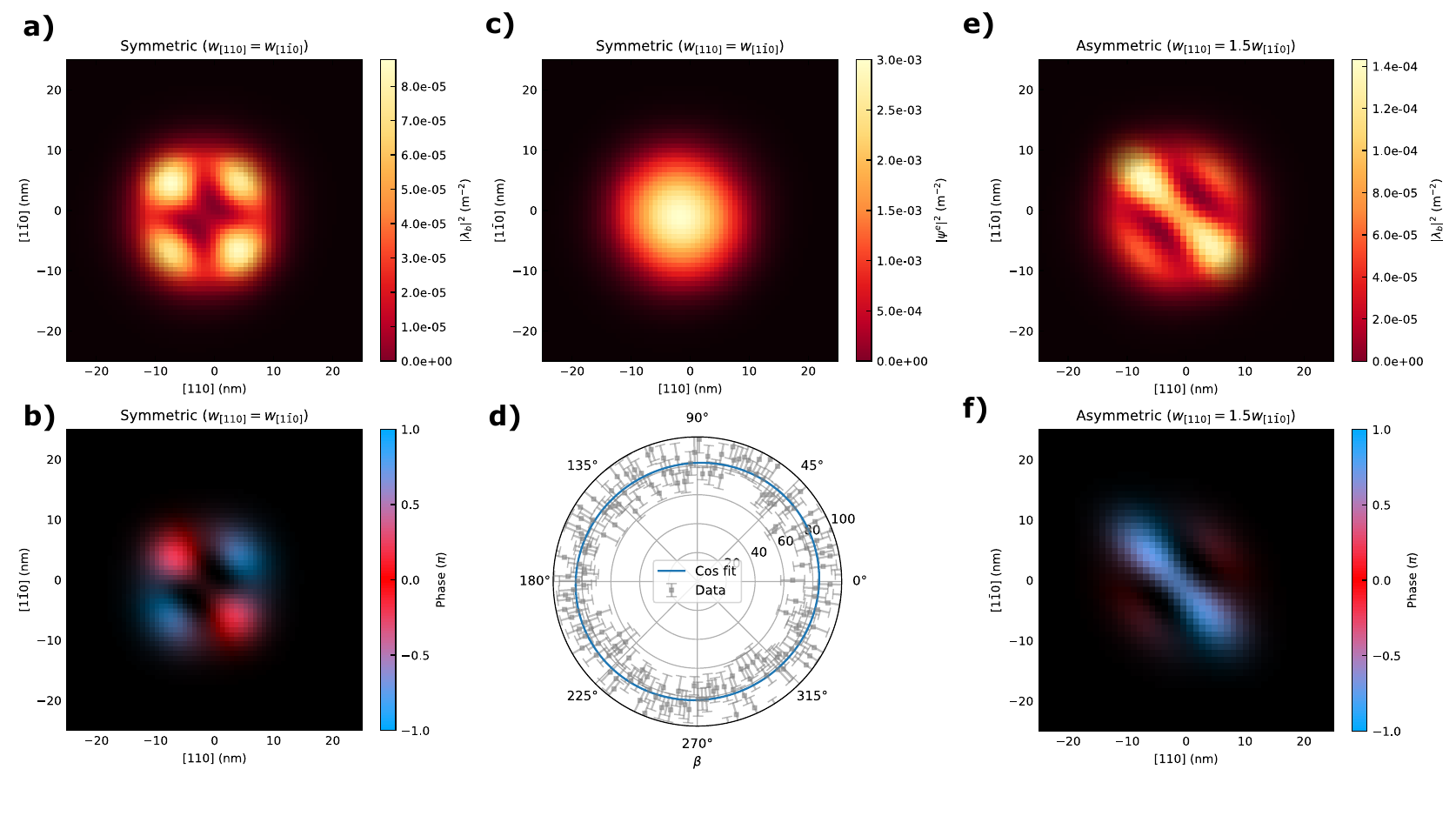}
    \caption{(a) Simulated spatial distribution of the envelope density $|\lambda_b(\bm{r})|^2$ of the $\ket{\Uparrow}$--$\ket{\downarrow}$ (``spin-bright'') LH admixture to the nominally HH wave function in the (001) plane (integrated along the [001] direction), for the case of an in-plane symmetric LDE GaAs QD (equal in-plane widths $w_{[1\bar{1}0]}\simeq w_{[110]}$). The density of the opposite $\ket{\Downarrow}$--$\ket{\uparrow}$ admixture is identical. (b) Complex phase of the LH wavefunction contribution, with the pixel brightnesses scaled by $|\lambda_b(\bm{r})|^2$. (c) Spatial distribution of the electron envelope function $|\psi^e(\bm{r})|^2$. (d) Intensity of the trion emission line at zero magnetic field as a function of the linear polarization angle $\beta$. The ratio between the two axes of elliptical polarization is $\epsilon\geq 0.92$. (e) Spatial distribution $|\lambda_b(\bm{r})|^2$ and (f) the complex phase for an asymmetric QD with $w_{[1\bar{1}0]}\simeq 1.5w_{[110]}$.} 
    \label{fig:MixingBright}
\end{figure*}

In Fig.~\ref{fig:MixingBright}, we evaluate the amount and impact of the $\ket{\Uparrow}$-$\ket{\downarrow}$ or $\ket{\Downarrow}$-$\ket{\uparrow}$ mixing in a LDE GaAs QD. This mixing can contribute to the HH-to-electron transition with equal energy and opposite circular polarization. Therefore, we refer to this kind of HH-LH mixing as ``spin-bright''. In this scenario, the HH-LH mixing can lead to a partially linear trion emission at zero magnetic field \cite{Belhadj2010} or at finite magnetic fields in the Faraday configuration. How much the emission is influenced, however, depends not only on the absolute degree of mixing but on the overlap of this admixture's envelope with the electron wave function, which ultimately determines its impact on the transition strength. Fig.~\ref{fig:MixingBright}(a) shows such an LH admixture envelope for a simulated LDE GaAs QD based on the same AFM nanohole scan as we use in the entire work. Summing over all pixels yields the total spin-bright contribution of $|\lambda_b|^2 \approx\SI{3.3}{\percent}$. Figure \ref{fig:MixingBright}(b) shows the corresponding complex phase distribution, with the brightness scaled by $|\lambda_b|^2$. The spatial dependence shows that the regions of strong mixing are concentrated close to the QD-barrier interface, where only tails of the electron wave function $|\psi^e(\bm{r})|^2$, shown in Fig. \ref{fig:MixingBright}(c), are located. Moreover, the admixture is predominantly composed of an odd contribution, as evidenced by Fig.~\ref{fig:MixingBright}(b), which further diminishes its overlap with the electron $s$-shell ground state. For this reason, even for the total $\ket{\Uparrow}$--$\ket{\downarrow}$/$\ket{\Downarrow}$--$\ket{\uparrow}$ mixing in unstrained LDE GaAs QDs reaching about \SI{4}{\percent}, the trion-electron transition polarization is still almost perfectly circular, as seen in Fig.~\ref{fig:MixingBright}(d), which shows an intensity measurement of a trion emission of one QD from the sample SA0952 as a function of the linear polarization angle $\beta$. The ratio between the two axes of elliptical polarization is $\epsilon\geq 0.92$, which yields a degree of circular polarization of $S^{(3)}/S^{(0)}=\SI{99.7}{\percent}$, with the deviation from unity mostly stemming from imperfections of the polarization analysis setup. The upper bound for the optically visible fraction of spin-bright HH-LH admixture is then given by $|\lambda_{b,\text{vis}}|^2 \lessapprox \frac{3(1-\sqrt{\epsilon})^2}{(1+\sqrt{\epsilon})^2} = \SI{0.1}{\percent}$ (see Ref. \cite{Belhadj2010}), which is far below the total spin-bright admixture of \SI{3.3}{\percent}. In the case of an in-plane asymmetry, as depicted in Figs.~\ref{fig:MixingBright}(e) and ~\ref{fig:MixingBright}(f), a new even contribution to the admixture arises, and as a result, the mixing-rich region overlapping with the electron increases, which leads to a stronger ellipticity of the polarization typical for InGaAs QDs. A similar effect is expected for in-plane uniaxial strain.

\begin{figure*}[h]
    \centering
    \includegraphics[]{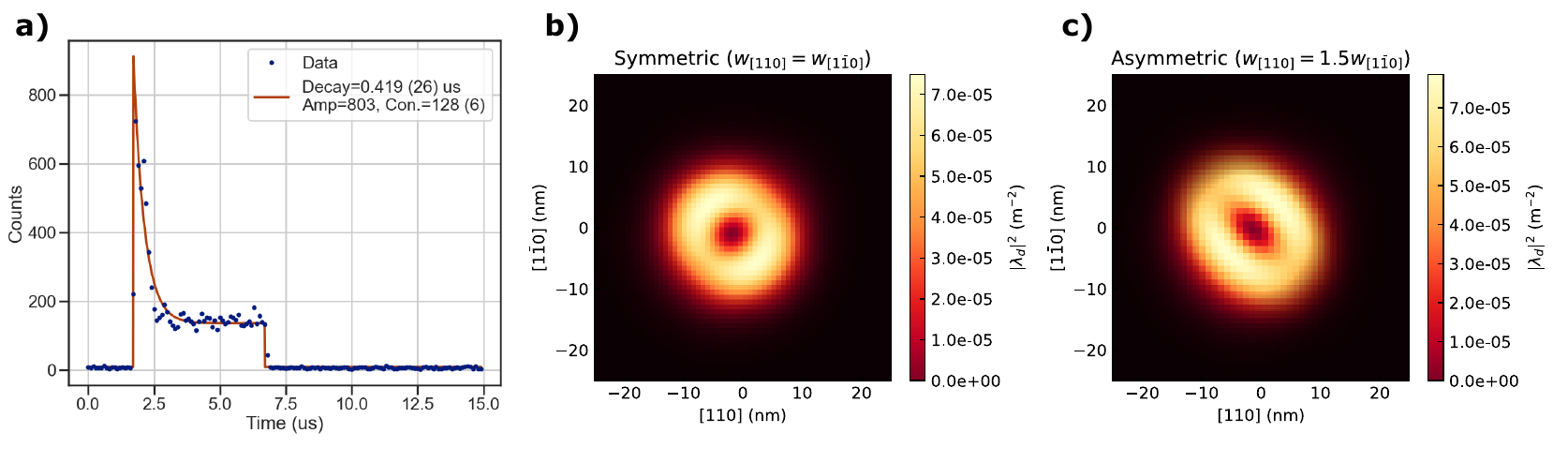}
    \caption{(a) Dynamics of a spin-pumping experiment in Faraday configuration on the sample Linz SA0952 with optical pulses generated by modulating a continuous wave, narrow-band laser with an acousto-optical modulator. (b) Simulated distribution of the envelope density $|\lambda_d(\bm{r})|^2$ of the $\ket{\Uparrow}$--$\ket{\uparrow}$ (``spin-dark'') LH admixture to the nominally HH ground state for an in-plane symmetric (equal in-plane widths $w_{[1\bar{1}0]}\simeq w_{[110]}$) and (c) an asymmetric QD ($w_{[1\bar{1}0]}\simeq 1.5w_{[110]}$).}
    \label{fig:MixingDark}
\end{figure*}

Figure~\ref{fig:MixingDark} treats a HH-LH mixing of nature $\ket{\Uparrow}$-$\ket{\uparrow}$ and $\ket{\Downarrow}$-$\ket{\downarrow}$, which we call the ``spin-dark`` mixing. This mixing is responsible for the finite transition into the opposite electron spin-state in optical-spin pumping in the Faraday configuration, where the diagonal transitions are not allowed in a pure HH system. Figure~\ref{fig:MixingDark}(a) shows the exponentially decaying scattering probability after switching on a CW laser in resonance with a HH-dipole-allowed transition. The excitation power was chosen as three times the saturation power of the two-level transition so that the spin-pumping rate is close to maximum. Comparing the decay time with the typical trion radiative lifetime of $T_1=\SI{230}{ps}$ \cite{Zhai2020} yields a cyclicity of $c=T_{\text{pump}}/T_1\lessapprox 2000$, i.e., the laser scatters $c$ times before the electron spin ends up in the opposite state and the driven transition becomes dark. The LH contributes with a factor of $1/\sqrt{3}$ to dipole transition compared to the HH, so we can roughly bound the fraction of the spin-dark mixing, which contributes to the optical electron-spin pumping, to $\sqrt{3}/\sqrt{c}\gtrapprox \SI{4}{\percent}$. The total spin-dark LH admixture to the hole ground state is calculated as about \SI{3.5}{\percent}, which means that a substantial amount of this admixture contributes to spin-pumping. Figure~\ref{fig:MixingDark}(b) shows the spatial distribution of the envelope density of the spin-dark LH admixture for the in-plane symmetric QD case and Fig.~\ref{fig:MixingDark}(c) for the asymmetric case. This mixing effect does not rely on the in-plane asymmetry, and, accordingly, the simulated admixture distribution, and thus its overlap with the electron wave function, changes little. Based on this, only a little change in spin-pumping behavior is expected in in-plane asymmetric QDs.

\subsection{Material parameters used in the simulations}

			\begingroup
			\begin{table}[]
				\newcommand{\cw}{\cite{Winkler2003}}
				\newcommand{\cs}{\cite{SaidiJAP2010}}
                \newcommand{\ccaro}{\cite{Caro2015}}
				\newcommand{\ca}{\cite{AmirtharajBOOK1994}}
                \newcommand{\cvlad}{\cite{Mlinar2005}}
                \newcommand{\clb}{\cite{Landolt-Bornstein}}
				\newcommand{\cmsq}{${\mathrm{C}}/{\mathrm{m}^2}$}
				\newcommand{\intEg}{$-0.13+1.31x$}
				\begin{tabular*}{0.5\textwidth}{@{\extracolsep{1cm}}l|lll}
					\toprule\rule{0pt}{1.1em} 
					& AlAs & GaAs & $C^\mathrm{AlAs}_{\mathrm{GaAs}}$ \\
					\hline\rule{0pt}{1.1em} 
					$a_\mathrm{}$\,(\AA) 			& 5.652	&5.642	&0\\
					$E_{\mathrm{g}}$\,(eV)			& 3.099	&1.519	&\intEg\\
					VBO\,(eV)						& -1.32	&-0.80	&0\\
					$E_{\mathrm{p}}$\,(eV)\cs$^{*}$ 	& 21.1	&28.8	&0\\
					$m_{\mathrm{e}}^{*}$ 				& 0.15	&0.0665	&0\\
					$\Delta_{\mathrm{SO}}$\,(eV) 		& 0.28	&0.341	&0\\
					$\gamma_{\mathrm{1}}$ 	           & 3.76	&6.98	&0\\
					$\gamma_{\mathrm{2}}$		           & 0.82	&2.06	&0\\
					$\gamma_{\mathrm{3}}$                 & 1.42	&2.93	&0\\
					$e_{\mathrm{14}}$\,(\cmsq)\ccaro 		& -0.055	&-0.205	&0\\ 
					$B_{\mathrm{114}}$\,(\cmsq)\ccaro 	& -1.61 	&-0.99	&0\\
					$B_{\mathrm{124}}$\,(\cmsq)\ccaro 	& -2.59 	&-3.21	&0\\
					$B_{\mathrm{156}}$\,(\cmsq)\ccaro 	& -1.32 	&-1.28	&0\\
					$C_{\mathrm{k}}$\,(eV\AA) 			& 0.002	&-0.0034	&0\\
					$a_{\mathrm{c}}$\,(eV)	 		& -5.64	&-7.17	&0\\
					$a_{\mathrm{v}}$\,(eV) 			& 2.47	&1.16	&0\\
					$b_{\mathrm{v}}$\,(eV) 			& -2.3	&-2.0	&0\\
					$d_{\mathrm{v}}$\,(eV) 			& -3.4	&-4.8	&0\\
					$c_{\mathrm{11}}$\,(GPa)			& 1250	&1211	&0\\
					$c_{\mathrm{12}}$\,(GPa) 			& 534	&566	&0\\
					$c_{\mathrm{44}}$\,(GPa) 			& 542	&600	&0\\
					$\varepsilon_{\mathrm{r}}$\cw 		&10.06	&12.4	&0\\
                    $g$\cw                         & 1.52  & $-0.44$  & 0\\
                    $\kappa$\cvlad                  & 0.12  & 1.28  & 0\\
                    $q$\clb                        & 0.04  & 0.04  & 0\\
					\toprule
                    \multicolumn{4}{p{0.5\columnwidth}}{
					$\!^{*}$Values used for calculation of optical properties; for the $\bm{k}\cdot\bm{p}$ Hamiltonian $E_{P} = \left({m_{0}}/{m_\mathrm{e}^{*}} - 1 \right) {E_\mathrm{g}(E_\mathrm{g}+\Delta)}/{(E_\mathrm{g}+2\Delta/3)}$ was used to preserve ellipticity of the $\bm{k}\cdot\bm{p}$ equation system for envelope functions \cite{VeprekPRB2007,BirnerBOOK2014}.}\\
				\end{tabular*}
				\caption{\label{tab:params} Material parameters used in the modeling of QDs and calculation of single-particle and exciton states; $C^\mathrm{A}_{\mathrm{B}}$ are values of ternary bowing parameters. Unless otherwise marked, parameters are taken after Ref.\,[\onlinecite{Vurgaftman2001}].}
			\end{table}
			\endgroup

The explicit form of the $\bm{k}\cdot\bm{p}$ Hamiltonian used can be found in Ref.~\cite{Mielnik2018}, while Table~\ref{tab:params} lists all the material parameters for GaAs and AlAs used in our simulations. For AlGaAs alloys, we use linear interpolation with possible additional bowing where given. The parameters come from Ref.~\cite{Vurgaftman2001} unless otherwise noted.

\section{Addtional information to optical measurements}

\subsection{Charge-tunable quantum dot devices}

\begin{figure*}[h]
    \centering
    \includegraphics[]{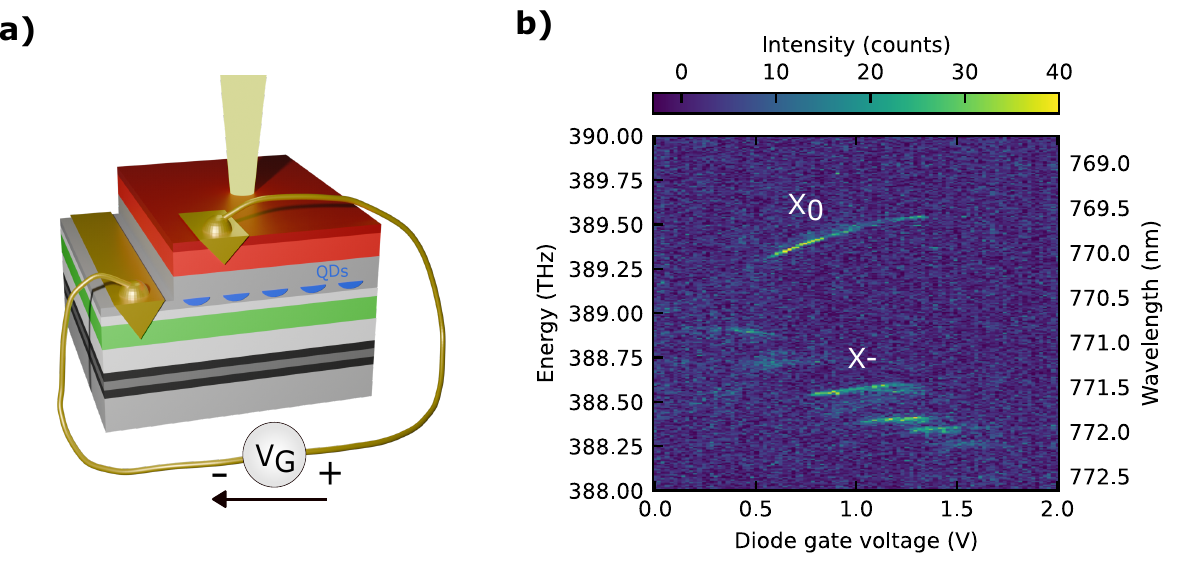}
    \caption{(a) Gated device used for the example QD in the main text. (b) PL spectra of a QD under above-bandgap excitation with CW laser at a wavelength of \SI{652}{nm} for varying gate voltage $V_\text{G}$. Around $V_\text{G}=\SI{1}{V}$, a single electron is likely to occupy the QD, which enables the trion (X$^-$) transition.}
    \label{fig:Device}
\end{figure*}

Most QDs used in this work (except Quandela AG170) are embedded in a p-i-n diode structure depicted in Fig.~\ref{fig:Device}(a), similar to the one used in Ref.~\cite{Zaporski2023}. Details about all the sample structures can be found in Sec.~\ref{sec:samples}. Figure~\ref{fig:Device}(b) shows the micro-photoluminescence spectra of an exemplary QD from sample Linz SA0952 with varying diode gate voltage $V_G$, excited by a CW laser at \SI{635}{nm} (well above the AlGaAs bandgap) in a confocal optical setup. In the range of $V_G=\SI{0.8}{V}$ to $\SI{1.3}{V}$ the QD is charged with a single electron, which allows the optical excitation of the negatively charged trion (X$^-$), as used in the main text.

\subsection{Estimation of the Stokes vectors from voltage-frequency scans}

\begin{figure*}[h]
    \centering
    \includegraphics[]{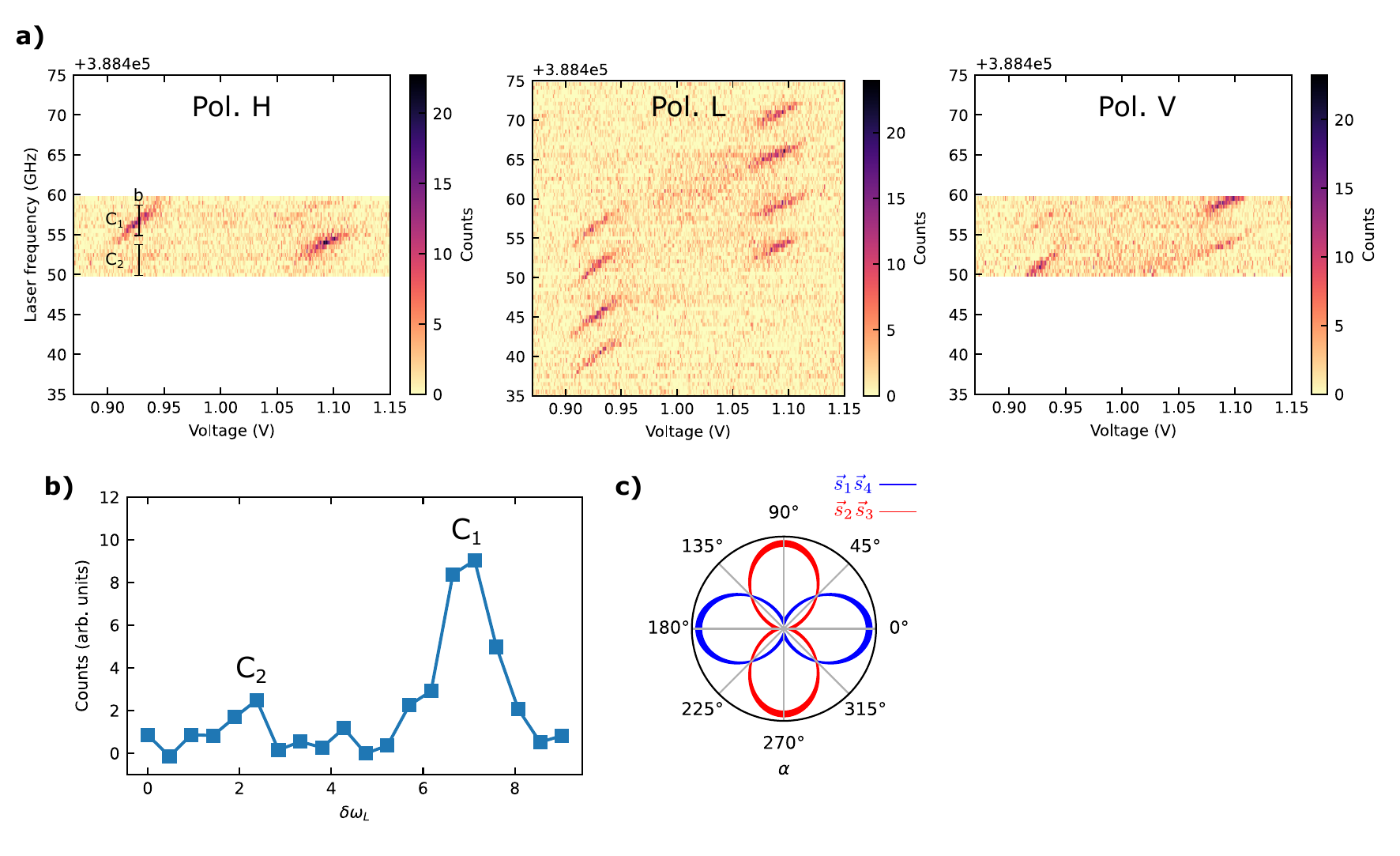}
    \caption{Polarization maps for the example QD from sample Linz SA0952 at $|\bm{B}|=\SI{5.8}{T}$ shown in the main text for horizontal ($H \parallel \bm{B}$), left circular (L) and vertical ($V \perp \bm{B}$) polarization. (b) The cross sections $C_1$ and $C_2$ at $V_G=\SI{0.93}{V}$ within a bin of $b=\SI{3.2} {mV}$ are used to determine $s_{1,4}^{(1)}=0.86(10)$ and $s_{2,3}=-0.86(10)$. (c) Linear polarization map with $\alpha$ relative to $\bm{B}$}
    \label{fig:PolMapsRF}
\end{figure*}

Figure~\ref{fig:PolMapsRF}(a) shows the polarization-resolved resonance fluorescence (RF) measurement from the same QD as in the main text but with the vertical ($V \perp \bm{B}$) polarization in addition to the left circular ($L$) and the horizontal ($H \parallel \bm{B}$)  polarizations. The $V$ polarization is, in principle, not necessary to estimate the degree of linear polarization and the alignment with respect to $\bm{B}$, but it demonstrates nicely that the opposite transitions $E_{1,4}$ vanish compared to the $E_{2,3}$ for a $H$-polarized laser. From these measurements, we cannot directly determine all Stokes vector components (this would require a full Stokes vector analysis), but we can set a lower bound on the rectilinear component $s^{(1)}$, which reflects the polarization along the $\bm{B}$ field (positive value) or perpendicular to it (negative value). We assume that we have negligible polarization scrambling in the system, i.e., the total intensity $s^{(0)}=\sqrt{s^{(1) 2} + s^{(2) 2} + s^{(3) 2}}$, with $s^{(2)}$ and $s^{(3)}$ being the diagonal and circular components, respectively, and the input laser arrives on the sample perfectly $H$ polarized. The latter was ensured in our setup by aligning the laser polarization to the plane of the optical table, which coincides with the magnetic field axis. In this configuration, the optical elements, of which the s- and p- polarization components align with the table frame, induce a minimum of ellipticity into the laser polarization. We can further assume from the nature of the four dipole moments that the dimmer transitions have the orthogonal polarization to the brighter ones. To estimate the relative strengths of the bright and dim transitions under a $H$-polarized laser, seen in Fig. \ref{fig:PolMapsRF}, we take the cross sections $C_1(\omega_L)$ and $C_2(\omega_L)$ at $V_G=\SI{0.93}{V}$ within a bin of $b=\SI{3.2} {mV}$ (8 pixels in this case), as seen in Fig. \ref{fig:PolMapsRF}(b). We can estimate the areas under the two curves (in arb. units) as $A_{C1}=17(4)$ and $A_{C2}=3(2)$, which results in $s_{1,4}^{(1)}=A_{C1}/(A_{C1}+A_{C2})=0.86(10)$, and the orthogonal components $s_{2,3}=-0.86(10)$. The linear polarization map for the two orthogonal polarization components is shown in Fig. \ref{fig:PolMapsRF}(c), with $\alpha$ the angle relative to $\bm{B}$.

\subsection{Estimation of the Stokes vectors from photoluminescence spectra}

\begin{figure*}[h]
    \centering
    \includegraphics[]{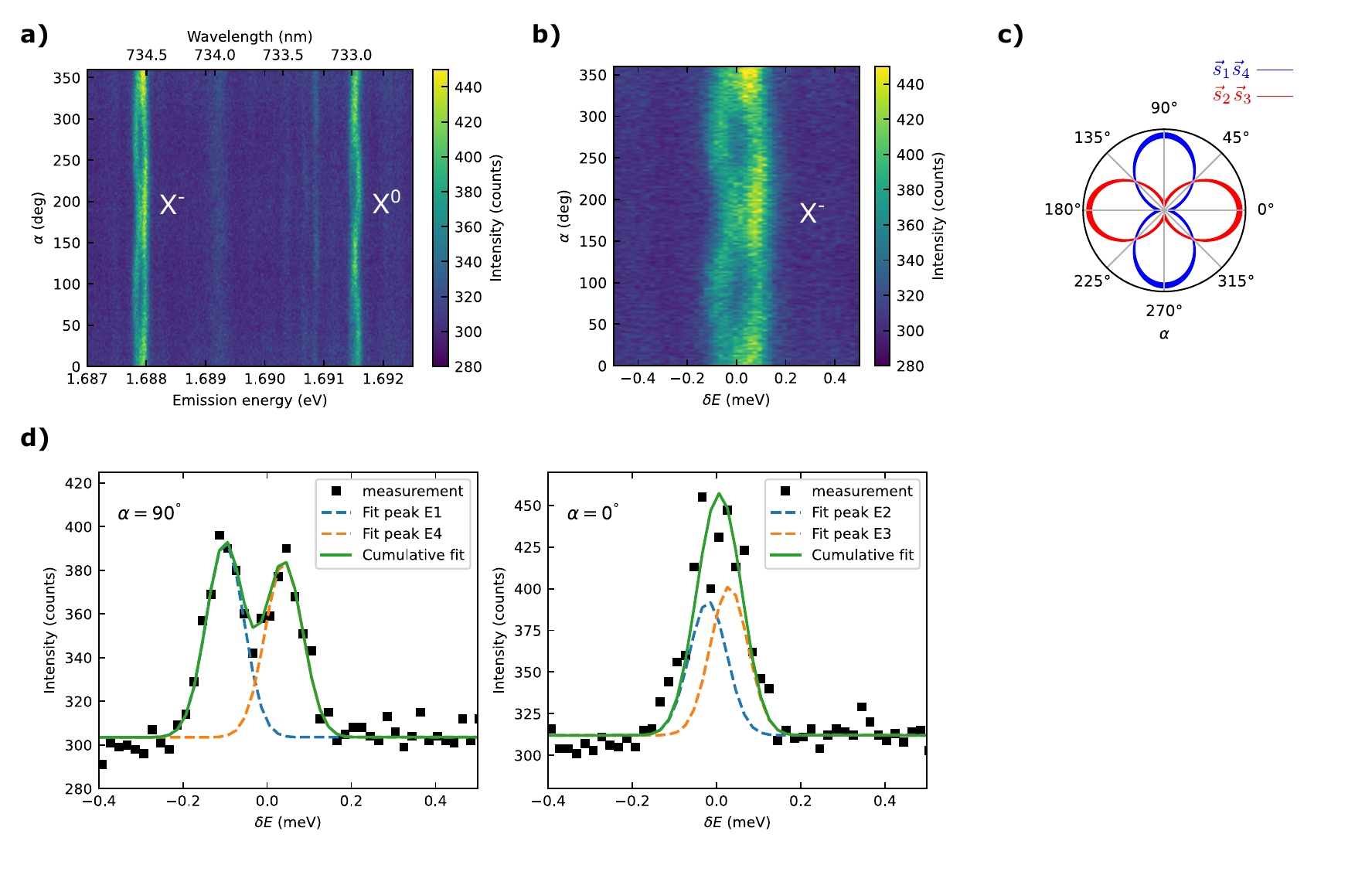}
    \caption{(a) Spectra of one QD from sample Quandela AG170 at $\chi=0$ and $\varphi=\pi/4$ and $|\bm{B}|=\SI{6}{T}$ as a function of $\alpha$, with the neutral exciton ($X^0$) and the trion ($X^-$) visible. (b) Zoom-in to the $X^-$ transition. (c) Alignment of the transition dipole moments deduced from (b). (c) Cross-sections taken from (b) for two different $\alpha$, from which $g^{e}=0.12(5)$ and $g^{t}=0.25(5)$ can be deduced by double Gaussian fits. The signs of the $g$-factors cannot be deduced by this method alone.}
    \label{fig:SpectraPolmaps}
\end{figure*}

If $|g^{e}|$ and $|g^{t}|$ are large and different enough to resolve all transitions $E_{1-4}$ sufficiently well with a spectrometer, both values can be found by a linear polarization map as shown in Fig.~\ref{fig:SpectraPolmaps} for a QD on Quandela AG170 sample at $\chi=\pi/2$ and $\varphi=\pi/4$ and $|\bm{B}|=\SI{6}{T}$. This sample is not gated (the QDs are not embedded in a diode), but still, the $X^-$ transitions are visible on average. Figure~\ref{fig:SpectraPolmaps}(a) shows a polarization map including the neutral exciton ($X^0$) and trion ($X^-$) emissions as a function of emission energy and linear polarization angle $\alpha$ relative to the in-plane component of the magnetic field $\bm{B}_{\perp}$. Figure~\ref{fig:SpectraPolmaps}(b) shows a zoom-in of the $X^-$. The resolution of the spectrometer is about $\SI{40}{\micro eV}$ ($\SI{10}{GHz}$), comparable to the observed Zeeman splitting at such field, so that the different emission lines as a function of $\alpha$ manifest in a continuous change of two Gaussian lines between two extreme cases: At $\alpha=\SI{90(5)}{\degree}$ $E_1$ and $E_4$ are mostly visible, and for $\alpha=\SI{0(5)}{\degree}$ $E_2$ and $E_3$ are dominating, which means that $\bm{s}_{1,4}\perp\bm{B}$ and $\bm{s}_{2,3}\perp\bm{B}$, as shown in Fig.~\ref{fig:PolMapsRF}(c). To find $|g^{e}|$ and $|g^{t}|$, we first look at a cross section at $\alpha=\SI{90}{\degree}$, shown in Fig.~\ref{fig:SpectraPolmaps}(d). Here we fit the data with a double Gaussian with equal standard deviations $\sigma=\SI{44(2)}{\micro eV}$ and the center energies for $E_1$ as $\omega_1=\SI{1.68758(1)}{\micro eV}$ and for $E_4$ as $\omega_4=\SI{1.68772(1)}{\micro eV}$. We then take the cross section at $\alpha=\SI{0}{\degree}$ and perform again a double Gaussian fit, assuming the same linewidth $\sigma=\SI{44(2)}{\micro eV}$ for both peaks. From there we deduce center energies for $E_2$ as $\omega_1=\SI{1.68766(1)}{\micro eV}$ and for $E_3$ as $\omega_4=\SI{1.68771(1)}{\micro eV}$. The Zeeman splittings for the electron and trion are then found by
\begin{equation}
    \begin{aligned}
    \omega_e &= \frac{\omega_2-\omega_1}{2} + \frac{\omega_4-\omega_3}{2}, \\
    \omega_t &= \frac{\omega_1+\omega_2}{2} - \frac{\omega_3+\omega_4}{2}. \\
    \end{aligned}    
\end{equation}
From these values, we can calculate the $g$-factors as
\begin{equation}
    g^{e,t} = \frac{\hbar\,\omega^{e,t}}{\mu_B\,|\bm{B}|},
\end{equation}
which in the case of this QD are $g^{e}=0.12(5)$ and $g^{t}=0.25(5)$. Note that the signs of the $g$-factors cannot be deduced directly from this measurement. In this case, the signs can safely be assumed to be positive based on the continuity of the curves in Fig.~5(b) in the main text.

\subsection{Influence of QD asymmetry on transition dipole moment orientation, indicated by exciton fine structure splitting}

\begin{figure*}[h]
    \centering
    \includegraphics[scale=1]{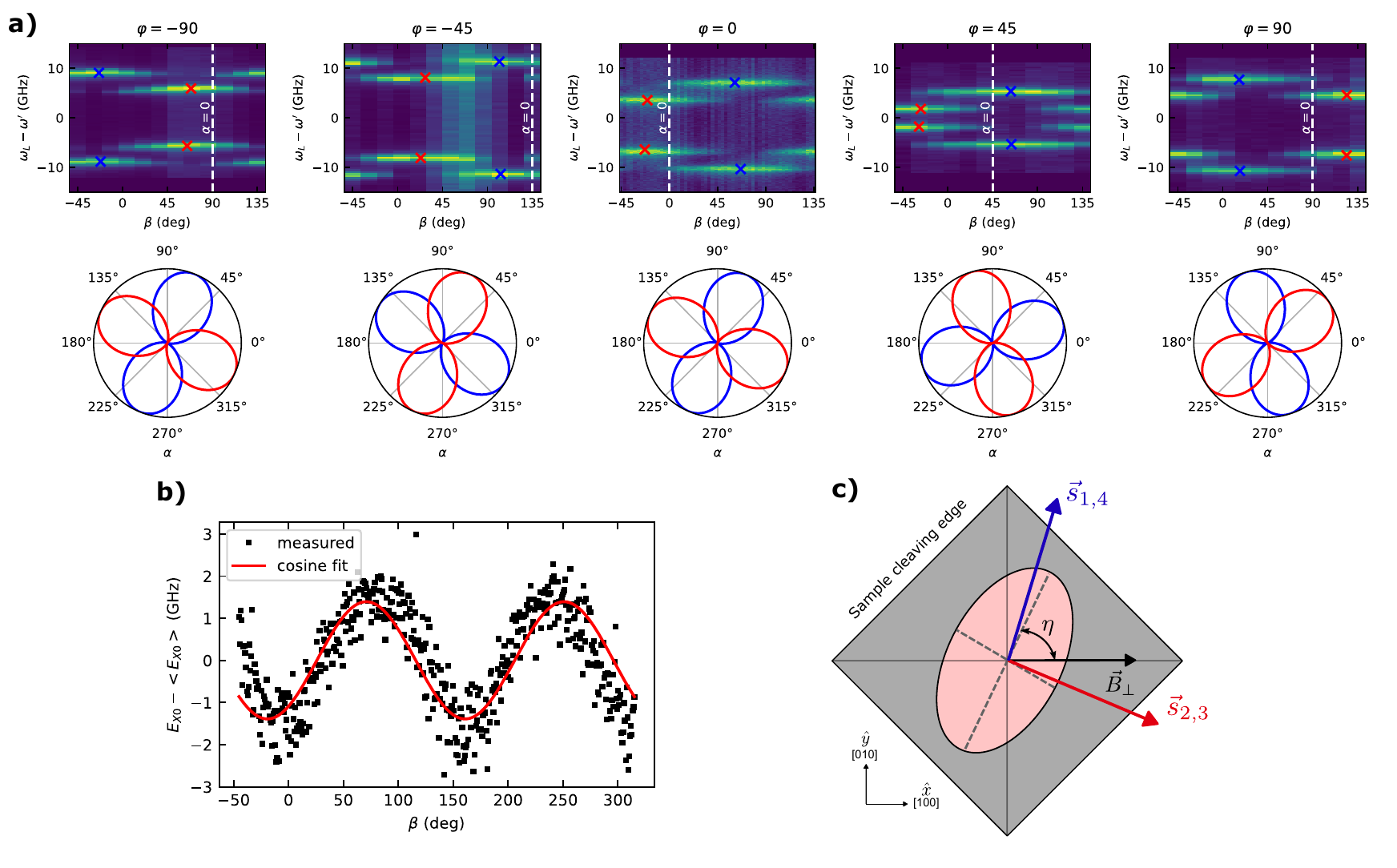}
    \caption{(a) Resonance fluorescence measurements of one QD from sample Basel 15446 emitting at $\omega^\prime=\SI{380.442}{THz}$ with $B_\perp=\SI{3}{T}$ under different $\varphi$. The polarization of the linear excitation laser $\beta$ is varied. The crosses mark the fitted $\beta$ at maximum intensity of $E_{1,4}$ (blue) and $E_{2,3}$ (red). The polar plots below show the TDM polarization and are obtained by a cosine fit of the intensity as a function of $\beta$ at a fixed $\omega_L$ at the resonances. (b) Peak position of the neutral exciton emission line as a function of linear polarization angle. The peak-to-peak amplitude of a cosine fit yields the fine structure splitting of \SI{3(1)}{GHz} (\SI{12(4)}{\micro eV}). The phase of the cosine fit yields the tilt of the asymmetry axis relative to the $\hat{x}$-axis of $\eta=\SI{71(1)}{\degree}$. (c) Sketch of a QD with asymmetric in-plane shape. The ellipsis depicts the anisotropy of the exchange interaction potential resulting in the FSS, with the highest value aligned at an angle $\eta$ relative to $\bm{B}$.}
    \label{fig:QDBasel}
\end{figure*}

Fine structure splitting (FSS) of the neutral excition in QDs arises from anisotropic in-plane electron-hole exchange interaction \cite{Plumhof2010}. In the case of an in-plane asymmetry of the QD (even in the absence of strain), the axes of the anisotropy follow the QD's elliptic shape. Figure~\ref{fig:QDBasel}(a) shows the RF measurements which were used to deduce the electron- and hole $g$-factor anisotropy in Fig.~5(c) in the main text. The polar plots show the fitted TDM alignments for all the different $\varphi$. In this measurement, $\varphi$ was changed by rotating the sample holder of a He bath cryostat by hand against a static magnet perpendicular to the optical axis. The precision of $\varphi$ is therefore limited to about $\pm\SI{5}{\degree}$, resulting in an error of the linear polarization of the TDMs of $\pm\SI{10}{\degree}$ according to Eq.~(4) in the main text. Even with this inaccuracy taken into account, we still observe a deviation from Eq.~(4), which cannot be explained by experimental errors. We believe that the deviation stems from the abnormally high FSS of $\SI{12(4)}{\micro eV}$ at an angle of $\eta=\SI{71(1)}{\degree}$ in this QD, which we deduce from Fig.~\ref{fig:QDBasel}(b). The high FSS suggests a high in-plane ellipticity of the QD, as sketched in Fig.~\ref{fig:QDBasel}(c), which partially pins the TDMs to the anisotropy-axis aligned at the angle $\eta$. However, the alternating alignment of the TDMs with every $\SI{45}{\degree}$ turn of $\varphi$ is still clearly visible in \ref{fig:QDBasel}(a), which is indicative for the dominating non-Zeeman Hamiltonian explained in Sec.~\ref{sec:TDM}.

\newpage
\section{GaAs QD samples}
\label{sec:samples}

\begin{figure}[h]
    \centering
    \includegraphics[]{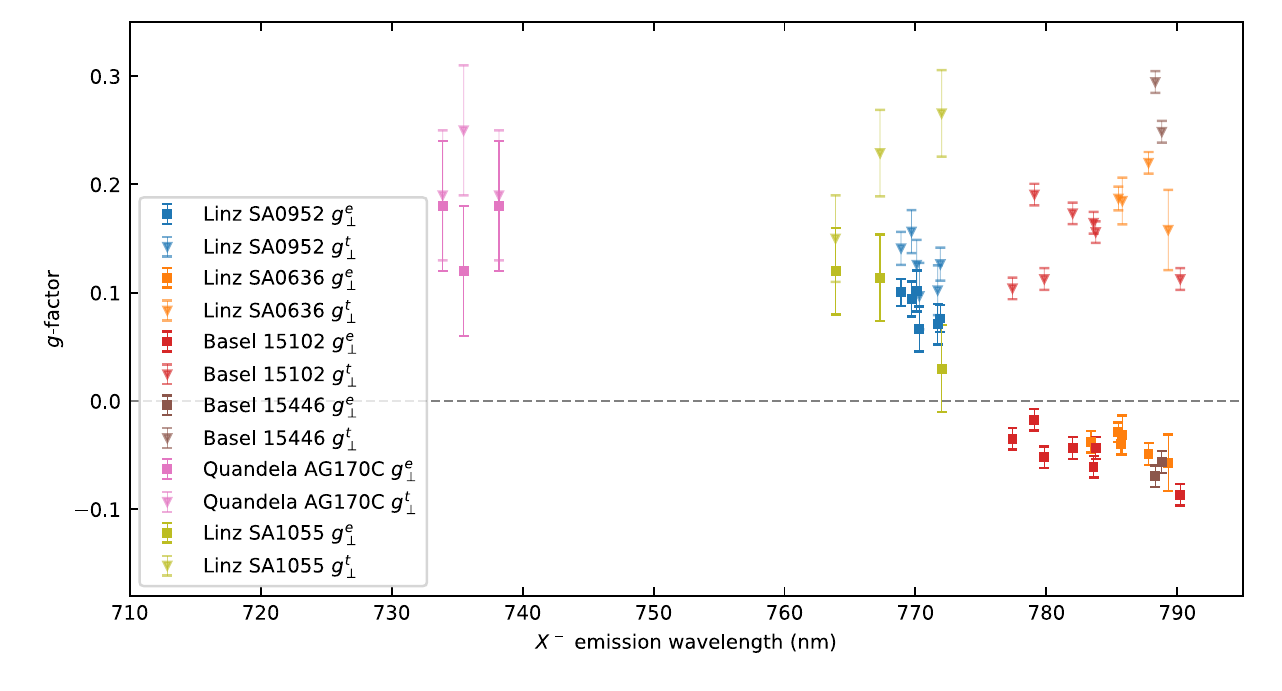}
    \caption{Overview of samples measured in this work}
    \label{fig:SamplesVoigt}
\end{figure}

\begin{table}[!ht]
  \centering
  \subfloat[Basel 15102]{
    \small
    \centering
      \begin{tabular}{c|c|c|c}
            Rep.   &thickness (nm) &material  & function \\
            \hline
            10x    &10.0  &GaAs &\\
            22x    &2.8   &AlAs & \\
                   &2.8   &GaAs &\\
            10x    &59.5  &Al0.33GaAs & DBR\\
                   &67.1  &AlAs &\\
            1x     &50.0  &Al0.15GaAs & \\
            1x     &150   &Al0.15GaAs & Si doped n-\\
            1x     &5.0   &Al0.15GaAs & segregation\\
            1x     &15.0  &Al0.15GaAs &\\
            1x     &10.0  &Al0.33GaAs & tunnel barrier\\
            1x     &0.4   &AlAs &\\
            1x     &2.0   &GaAs & filling (QDs)\\
            1x     &273.6 &Al0.33GaAs &\\
            1x     &65.0  &Al0.15GaAs & C doped p+\\
            1x     &10.0  &Al0.15GaAs & C doped p++\\
            1x     &5.0   &GaAs & C doped p++\\
            \end{tabular}
  }
  \hspace{1cm}
  \subfloat[Basel 15446]{
    \small
    \centering
    \begin{tabular}{c|c|c|c}
        Rep.   &thickness (nm) &material  & function \\
        \hline
        10x    &10.0  &GaAs &\\
        22x    &2.8   &AlAs & \\
               &2.8   &GaAs &\\
               &59.5  &Al0.33GaAs &\\
        9x     &67.1  &AlAs & DBR\\
               &59.5  &Al0.33GaAs &\\
        1x     &67.1  &AlAs\\
        1x     &50.0  &Al0.15GaAs & \\
        1x     &150   &Al0.15GaAs & Si doped n-\\
        1x     &5.0   &Al0.15GaAs & segregation\\
        1x     &15.0  &Al0.15GaAs &\\
        1x     &10.0  &Al0.33GaAs & tunnel barrier\\
        1x     &0.3   &AlAs &\\
        1x     &1     &GaAs & filling (QDs)\\
        1x     &273.6 &Al0.33GaAs &\\
        1x     &65.0  &Al0.15GaAs & C doped p+\\
        1x     &10.0  &Al0.15GaAs & C doped p++\\
        1x     &5.0   &GaAs & C doped p++\\
        \end{tabular}
  }
  \caption{Diode samples from Basel. ``Rep'' refers to the number of repetitions of the sub-structure.}
  \label{tab:SamplesBasel}
\end{table}

\begin{table}[!ht]
  \centering
  \subfloat[Linz SA0952]{
    \small
    \centering
         \begin{tabular}{c|c|c|c}
            Rep.   &thickness (nm) &material  & function \\
            \hline
            1x     &69.0  &Al0.95GaAs & DBR\\
                   &59.9  &Al0.2GaAs & \\
                   &69.0  &Al0.95GaAs &\\
            1x     &88.5  &Al0.15GaAs & \\
            1x     &100   &Al0.15GaAs & Si doped n-\\
            1x     &5.3   &Al0.15GaAs & segregation\\
            1x     &10.3  &Al0.15GaAs &\\
            1x     &18.8  &Al0.33GaAs & tunnel barrier\\
            1x     &0.8   &GaAs & filling (QDs)\\
            1x     &261.4 &Al0.33GaAs &\\
            1x     &68.3  &Al0.15GaAs & C doped p+\\
            1x     &5.3   &Al0.15GaAs & C doped p++\\
            1x     &10.5  &GaAs & C doped p++\\
            \end{tabular}
  }
  \hspace{1cm}
  \subfloat[Linz SA0636]{
    \small
    \centering
        \begin{tabular}{c|c|c|c}
            Rep.   &thickness (nm) &material  & function \\
            \hline
            1x     &69.0  &Al0.95GaAs & DBR\\
            6x     &59.9  &Al0.2GaAs & \\
                   &69.0  &Al0.95GaAs &\\
            1x     &93.5  &Al0.15GaAs & \\
            1x     &100   &Al0.15GaAs & Si doped n-\\
            1x     &5.3   &Al0.15GaAs & segregation\\
            1x     &10.5  &Al0.15GaAs &\\
            1x     &22.1  &Al0.33GaAs & tunnel barrier\\
            1x     &1.8   &GaAs & filling (QDs)\\
            1x     &281.4 &Al0.33GaAs &\\
            1x     &68.3  &Al0.15GaAs & C doped p+\\
            1x     &5.3   &Al0.15GaAs & C doped p++\\
            1x     &10.5  &GaAs & C doped p++\\
            \end{tabular}
  }
    \hspace{1cm}
    \subfloat[Linz SA1055]{
    \small
    \centering
            \begin{tabular}{c|c|c|c}
            Rep.   &thickness (nm) &material  & function \\
            \hline
            5x     &66.0  &AlAs & DBR\\
                   &55.6  &Al0.15GaAs & \\
            1x     &66.0  &AlGaAs &\\
            1x     &88.5  &Al0.15GaAs & \\
            1x     &100   &Al0.15GaAs & Si doped n-\\
            1x     &5.3   &Al0.15GaAs & segregation\\
            1x     &10.3  &Al0.15GaAs &\\
            1x     &18.8  &Al0.33GaAs & tunnel barrier\\
            1x     &0.8   &GaAs & filling (QDs)\\
            1x     &261.4 &Al0.33GaAs &\\
            1x     &73.5  &Al0.15GaAs & C doped p+\\
            1x     &66.0  &AlAs & DBR\\
                   &55.6  &Al0.15GaAs & \\
            \end{tabular}
  }
  \caption{Diode samples from Linz. ``Rep'' refers to the number of repetitions of the sub-structure.}
  \label{tab:SamplesLinz}
\end{table}

\begin{table}[!ht]
    \centering
    \begin{tabular}{c|c|c|c}
    Rep.   &thickness (nm) &material & function\\
    \hline
    20x    &59.9  &Al0.15GaAs & DBR\\
           &65.2  &Al0.95GaAs &\\
    1x     &116.5 &Al0.33GaAs &\\
    1x     &3     &GaAs & QDs\\
    1x     &116.5 &Al0.33GaAs &\\
    10x    &59.9  &Al0.95GaAs & DBR\\ 
           &65.2  &Al0.10GaAs &\\
    \end{tabular}
    \caption{Quandela AG170 - not gated, ``Rep'' refers to the number of repetitions of the sub-structure.}
    \label{tab:my_label}
\end{table}
\FloatBarrier
\newpage
\bibliography{supplement}